\documentclass[useAMS,usenatbib]{mn2e}

\usepackage{times,verbatim,graphicx}
\usepackage{url,epsfig,dcolumn,amsmath}
\usepackage{color}
\definecolor{red}{rgb}{0.75,0.0,0.0}
\definecolor{green}{rgb}{0.0,0.75,0.0}
\definecolor{blue}{rgb}{0.0,0.0,0.75}
\newcolumntype{.}{D{.}{.}{4}}
\newcolumntype{,}{D{.}{.}{2}}
\newcolumntype{;}{D{.}{.}{1}}
\newcommand{\exels}{ExELS}
\newcommand{\euclid}{\emph{Euclid}}
\newcommand{\kepler}{\emph{Kepler}}
\newcommand{\rjup}{\mbox{R}_{\rm Jup}}
\newcommand{\mjup}{\mbox{M}_{\rm Jup}}

\newcommand{\lesssim}{{\lower-1.2pt\vbox{\hbox{\rlap{$<$}\lower5pt\vbox{\hbox{$\sim$}}}}}}
\newcommand{\gtrsim}{{\lower-1.2pt\vbox{\hbox{\rlap{$>$}\lower5pt\vbox{\hbox{$\sim$}}}}}}
\newcommand{\besancon}{Besan{\c c}on}
\newcommand{\vis}{{\sc VIS}}
\newcommand{\nisp}{{\sc NISP}}

\title[\exels{} II. Hot exoplanets and sub-stellar systems]{\exels{}: an exoplanet legacy science proposal for the ESA \euclid{} mission II. Hot exoplanets and sub-stellar systems}
\author[I. McDonald et al.]{I.~McDonald$^{1,2}$\thanks{E-mail:
mcdonald@jb.man.ac.uk}, E. Kerins$^{1,2}$, M. Penny,$^{1,2,3}$
  J.-P.~Beaulieu,$^{1,4}$ V.~Batista,$^{1,3}$ 
\newauthor
  S.~Calchi~Novati,$^{1,7,8,9}$ A.~Cassan,$^{1,4}$ P.~Fouqu\'{e},$^{1,10}$ S.~Mao,$^{1,2,6}$ J.B.~Marquette,$^{1,4}$ 
\newauthor
   N.~Rattenbury,$^2$  A.C.~Robin,$^5$ P.~Tisserand,$^{1,11}$ M.R.~Zapatero~Osorio$^{1,12}$\\
$^1$ The Euclid Exoplanet Science Working Group\\
$^2$Jodrell Bank Centre for Astrophysics, School of Physics \& Astronomy, University of Manchester, Oxford Road, Manchester M13 9PL, UK\\
$^3$Department of Astronomy, Ohio State University, 140 W. 18th Ave., Columbus, OH 43210, USA\\
$^4$Institut d`Astrophysique de Paris, Universit\'{e} Pierre et Marie Curie, CNRS UMR7095, 98bis Boulevard Arago, 75014 Paris, France\\
$^5$Institut Utinam, CNRS UMR6213, Universit\'e de Franche-Comt\'e, OSU THETA Franche-Comt\'e-Bourgogne, Besan\c{c}on, France\\
$^6$National Astronomical Observatories, Chinese Academy of Sciences, A20 Datun Road, Chaoyang District, Beijing 100012, China\\
$^7$Dipartimento di Fisica ``E. R. Caianiello'', Universit\`a di Salerno, Via Ponte don Melillo, 84084 Fisciano (SA), Italy\\
$^8$Istituto Internazionale per gli Alti Studi Scientifici (IIASS), Vietri Sul Mare (SA), Italy\\
$^9$NASA Exoplanet Science Institute, MS 100-22, California Institute of Technology, Pasadena CA 91125 (Sagan visiting fellow)\\
$^{10}$IRAP, CNRS - Université de Toulouse, 14 av. E. Belin, F-31400 Toulouse, France \\
$^{11}$Research School of Astronomy and Astrophysics, Australian National University, Cotter Rd, Weston Creek, ACT, 2611, Australia\\
$^{12}$Centro de Astrobiolog\'{i}a (CSIC-INTA), Crta. Ajalvir km 4, E-28850 Torrej\'{o}n de Ardoz, Madrid, Spain
}

\begin{document}

\date{Accepted 9999 December 32. Received 9999 December 32; in original form 9999 December 32}

\pagerange{\pageref{firstpage}--\pageref{lastpage}} \pubyear{9999}

\maketitle

\label{firstpage}

\begin{abstract}
The Exoplanet \euclid{} Legacy Survey (\exels{}) proposes to determine the frequency of cold exoplanets down to Earth mass from host separations of $\sim$1~AU out to the free-floating regime by detecting microlensing events in Galactic Bulge. We show that \exels{} can also detect large numbers of hot, transiting exoplanets in the same population. The combined microlensing+transit survey would allow the first self-consistent estimate of the relative frequencies of hot and cold sub-stellar companions, reducing biases in comparing ``near-field'' radial velocity and transiting exoplanets with ``far-field'' microlensing exoplanets. The age of the Bulge and its spread in metallicity further allows \exels{} to better constrain both the variation of companion frequency with metallicity and statistically explore the strength of star--planet tides.

We conservatively estimate that \exels{} will detect $\sim$4\,100 sub-stellar objects, with sensitivity typically reaching down to Neptune-mass planets. Of these, $\sim$600 will be detectable in both \emph{Euclid's} \vis{} (optical) channel and \nisp{} $H$-band imager, with $\sim$90 per cent of detections being hot Jupiters. Likely scenarios predict a range of 2\,900\ --\ 7\,000 for \vis{} and 400\ --\ 1\,600 for $H$-band. Twice as many can be expected in \vis{} if the cadence can be increased to match the 20-minute $H$-band cadence. The separation of planets from brown dwarfs via Doppler boosting or ellipsoidal variability will be possible in a handful of cases. Radial velocity confirmation should be possible in some cases, using 30-metre-class telescopes. We expect secondary eclipses, and reflection and emission from planets to be detectable in up to $\sim$100 systems in both \vis{} and \nisp{}-$H$. Transits of $\sim$500 planetary-radius companions will be characterised with two-colour photometry and $\sim$40 with four-colour photometry (\vis{},$YJH$), and the albedo of (and emission from) a large sample of hot Jupiters in the $H$-band can be explored statistically.
\end{abstract}

\begin{keywords}
planetary systems --- planets and satellites: detection --- stars: variables: general --- Galaxy: bulge --- stars: low-mass --- planet--star interactions
\end{keywords}


\section{Introduction}
\label{IntroSect}

Nearly 1800 exoplanets are now known\footnote{As of June 2014; see: \url{http://exoplanet.eu/}}, primarily from the radial velocity and transit methods, and another $\sim 1700$ candidates from transit observations with the \kepler{} space telescope \citep{BRB+12} still await confirmation. Both techniques are most sensitive to more massive planets and/or planets close to their host star ($\lesssim$1 AU).

Detection of planets further from their star typically relies on either direct imaging, which only works for the closest and youngest systems (e.g.\ \citealt{CLD04,KGC+08,MMB+08}), or gravitational microlensing for systems much further away (e.g.\ \citealt{BUJ+04,BBF+06}). While less prolific (accounting for $\approx$4 per cent of detected planets), these techniques are nevertheless the most productive means of identifying planets beyond $\sim$1 AU.

A number of statistical studies using different techniques have recently derived first robust constraints on the frequency of exoplanets for different ranges of masses and orbital separations (e.g., \citealt{CBM+08,GDG+10,MML+11,BKB+11a,CKB+12,HMB+12,FTC+13}). A study combining results from all techniques would further calibrate the planet frequency across a range of masses and separations from their host star (cf.\ \citealt{QLM+12,CG14a,CG14}). As all four techniques lose sensitivity near the habitable zone of solar-like stars, accurately determining the number of habitable, Earth-like planets in the Galaxy requires a combination of different techniques. Such comparisons are not presently straightforward: microlensing typically detects planets at distances of several kiloparsecs from the observer, whereas the radial velocity and transit methods mostly focus on brighter stars within 1~kpc of the Sun, and direct imaging is only effective within a few tens of parsecs. Statistical studies are therefore limited by systematic variation of parameters such as host star metallicity and age, and are limited by the small number of planets detected at large radii from their hosts.

\subsection{The \exels{} observing strategy} \label{exels}

\euclid{} \citep{LAA+11} will comprise a 1.2-m Korsch telescope imaged by both optical and near-infrared cameras via a dichroic splitter. The optical camera (\vis{}) will use a single, broad, unfiltered bandpass, spanning roughly the $RIZ$ bands, and a mosaic of $36 \times 4{\rm k} \times 4{\rm k}$ CCDs with 0.1~arcsec pixels. The near-infrared camera (\nisp{}) will have three broadband filters ($Y$, $J$ and $H$ bands) with a mosaic of $16 \times 2{\rm k} \times 2{\rm k}$ HgCdTe arrays with 0.3~arcsec pixels, giving a total field of view of $\approx$0.54 deg$^2$. 

\euclid{} is designed to study dark energy through weak lensing and baryon acoustic oscillations \citep{LAA+11}. However it is also expected to undertake some additional science programmes. We have previous proposed the Exoplanet \euclid{} Legacy Survey (\exels{}; \citealt{PKR+13}; hereafter Paper~I). \exels{} would use the gravitational microlensing method to perform the first space-based survey of cold exoplanets (typically $>$1~AU from their hosts, including ``free-floating'' planets).

\exels{} proposes sustained high-cadence observations of three adjacent fields (1.64~deg$^2$) towards the Galactic Bulge, with a planned centre at $l = 1.1^{\circ}, b = -1.7^{\circ}$. These fields will be observed continuously for periods of up to one month, limited by \emph{Euclid's} solar aspect angle constraint, with a cadence of 20~minutes in the \nisp{}-$H$ channel. The \euclid{} on-board storage capacity and downlink rate of 850~Gb/day should also allow observations in the \vis{} channel with a cadence as short as one hour. The \vis{} data (in theory, comprising of stacks of several exposures taken over each hour) can provide the source--lens relative proper motion for many microlensing events and, as we show, would provide the primary channel for detecting transiting systems. The theoretical maximum (adopted both in Paper~I and here) of 10 $\times$ 30-day observation blocks is unlikely to be possible within the main 6-year \euclid{} mission lifetime, but could be completed after the primary cosmology mission if the spacecraft remains in good health.

\exels{} primary science objective is the detection of cool, low-mass exoplanets using microlensing. As such the observing strategy is determined by and optimised for microlensing rather than for transit observations. The observing strategy is therefore taken to be fixed and accepted as non-optimal for the transit science we discuss in this paper. The Galactic Centre is not an ideal region to look for transiting exoplanets due to severe crowding. Nonetheless, it is a viable region: both the ground-based Optical Gravitational Lensing Experiment (OGLE) microlensing survey \citep[e.g.][]{UPN+08} and the dedicated \emph{Hubble Space Telescope} Sagittarius Window Eclipsing Extrasolar Planet Search (SWEEPS) survey \citep{SCB+06} have detected transiting planets here. \citet{BR02} predict that $\sim$50\,000 transiting planets could be found using a survey very similar to that proposed by \exels{}, though without the dedicated, continuous coverage of their proposal, we cannot expect quite so many here.

\exels{} will provide a more distant sample of hot, transiting exoplanets than obtained from the current and planned large-scale dedicated transit surveys. Yet we will show that a space-based survey should produce many more transits than those detected by Bulge-staring surveys like OGLE and SWEEPS. The combination should be especially useful for removing environmental biases in comparisons of more distant microlensing samples to the nearby transit samples from \kepler{}, \emph{PLATO}, the \emph{Transiting Exoplanet Survey Satellite} (\emph{TESS}) and ground-based transit surveys.

The remainder of this paper is organised as follows. In Section~\ref{ModelSect}, we describe the modelled stellar population, simulated observations, and noise estimation. Section~\ref{SNSect} details the mathematical model determining which planets are detectable. Section~\ref{ExoSect} describes the simulation of the underlying exoplanet population, and the physical and technical uncertainties in planet detectability. In Section~\ref{ResultsSect}, we detail our results, the number and properties of detectable planets and their hosts, and their uncertainties, and compare them with those of the associated microlensing survey. We also examine follow-up strategies for detected planets.

\section{Transit simulations}
\label{ModelSect}

\subsection{Terminology}
\label{ModelTermSect}

\begin{figure}
\centerline{\includegraphics[width=0.33\textwidth,angle=0]{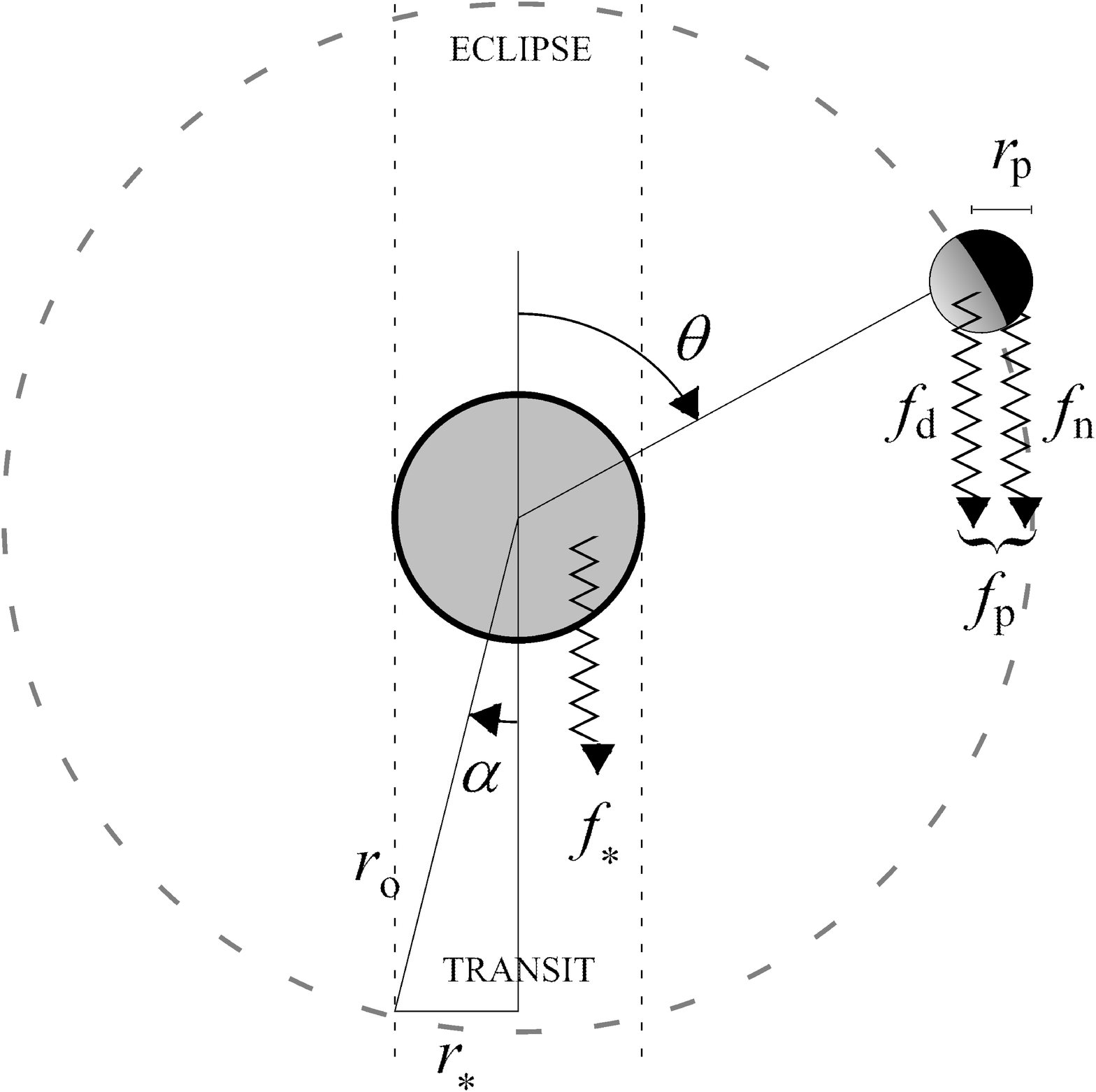}}
\centerline{\includegraphics[width=0.33\textwidth,angle=0]{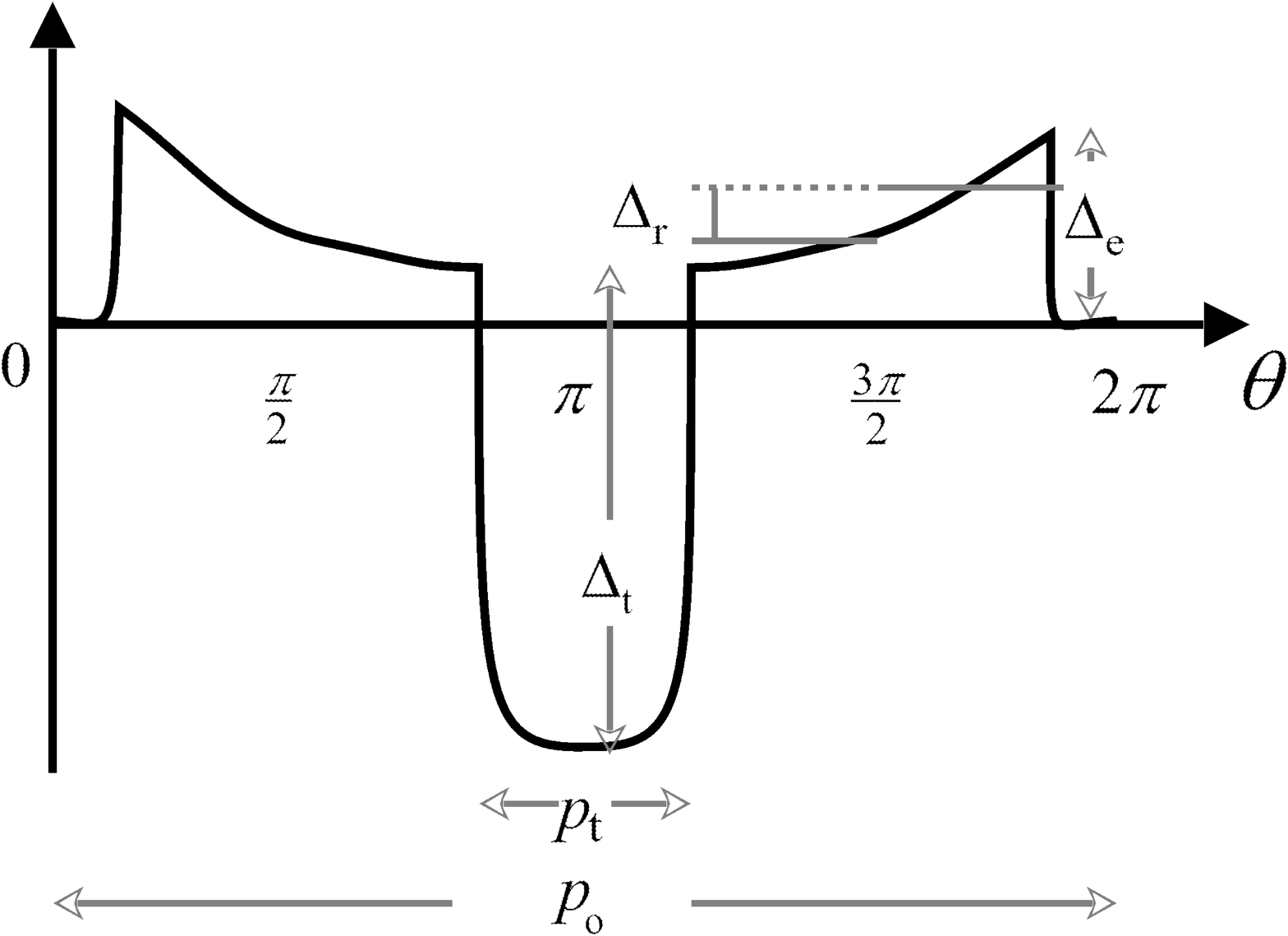}}
\caption{Schematic diagrams of an orbiting planet (top) and the transit light curve (bottom) showing terms used in this work, including the planetary ($r_{\rm p}$), stellar ($r_\ast$) and orbital ($r_{\rm o}$) radii; the stellar ($f_\ast$) and planetary flux ($f_{\rm p}$, split into day and night hemispheres as $f_{\rm d}$ and $f_{\rm n}$); the phase angle ($\theta$) and transit duration angle ($\alpha$); the orbital ($p_{\rm o}$) and transit ($p_{\rm t}$) periods; and the amplitude of lightcurve modulations caused by transit ($\Delta_{\rm t}$), eclipse ($\Delta_{\rm e}$) and reflection ($\Delta_{\rm r}$) of and from the planet.}
\label{SchemFig}
\end{figure}

We will consider simple binary systems with stellar-mass primary host stars and low-mass secondary companions. As this is a transit survey, we must describe companions based on their radius ($r_{\rm p}$), as well as their mass ($m_{\rm p}$), which we do as follows for mass:
\begin{description}
\item[\it Planets:] Objects with $m_{\rm p} < 13~\mjup$;
\item[\it Brown dwarfs:] Objects with $13 \leq m_{\rm p} < 75~\mjup$;
\item[\it Low-mass stars:] Objects with $m_{\rm p} \geq 75~\mjup$;
\end{description}
and for radius:
\begin{description}
\item[\it Small planets:] Objects with $r_{\rm p} < 0.7~\rjup$;
\item[\it Jupiter-radius objects:] Objects with $0.7~\rjup \leq r_{\rm p} \leq 1.4~\rjup$;
\item[\it Small stars:] Objects with $1.4~\rjup < r_{\rm p} < 3~\rjup$.
\end{description}
Figure \ref{SchemFig} shows a schematic defining many of the symbols used in the following sections.

\subsection{Survey assumptions and model inputs}
\label{ModelGalSect}

Our simulated survey mirrors the assumptions of the \exels{} microlensing survey described in Paper~I, but with three alterations.

{\it Observing cadence:} The \vis{} cadence is increased from every 12 hours to every hour (Section~\ref{exels}). Planets will be primarily detected in the \vis{} channel (Section \ref{ResultsSect}), making this a requirement of a transit survey. The number of planets found in the microlensing survey would be slightly increased from the values in Paper~I, while colour data from lensing events would constrain properties of many more lower-mass free-floating planets. One hour is currently the maximum allowed by the agreed telemetry rate.

{\it Secondary masses:} We consider an upper limit of 0.3~M$_\odot$, rather than 0.03~M$_{\odot}$. The flatness of the mass-radius relation for large planets, brown dwarfs and low-mass stars means that transits of large Jupiter-like planets cannot be distinguished from those of brown-dwarf or low-mass stellar binaries.

{\it Planet frequency:} This is scaled to match that of observed `hot' exoplanet frequency (see Section~\ref{ModelPlanOrbitSect}).

We model our exoplanet host population using a Monte-Carlo approach, based on version 1106 of the \besancon\ population synthesis Galaxy model \citep{RRDP03}, which incorporates a three-dimensional dust model to compute extinction and reddening \citep{MRRS05}. The populations included in the model are such that stars are limited to between 0.08 -- 4 M$_\odot$ (B--M type stars). No transiting planets have yet been found around stars outside this range and the proposed target area does not include large or young open clusters or star-forming regions. Evolution is covered from the main sequence to a bolometric luminosity of $M_{\rm bol} \approx 0$ on the giant branch. The lack of evolved giants, which can be significantly variable at the wavelengths considered (e.g.\ \citealt{MZS+14}), is expected to be an unmodelled source of red noise and false positives in our sample.

Simulated images were constructed from stars drawn from the \besancon\ model, with a stellar density equivalent to 29~million stars with $m_{VIS} < 24$, or 108~million stars with $m_{\rm H} < 24$, over the 1.64 square degree survey area. Each star is simulated using a numerical point-spread function (PSF) computed for the appropriate filter and detector\footnote{PSFs come from M.\ Cropper and G.\ Seidel, private communications, and include the effects of telescope jitter and telescope and instrument optics in different bands in an identical fashion to Paper~I.}. To obtain a conservative estimate, PSFs were computed near the edge of the detector field of view, and with throughput computed for sub-optimal detector performance to include some of the effects of aging on the CCD. Full details of these simulations are given in Paper~I.

Aperture photometry was performed on these simulated images to extract photometric magnitudes. This provides a conservative estimate for photometric uncertainty due to blending in crowded fields, where PSF-fitting or difference imaging would be more accurate. Apertures of radius 0.8$^{\prime\prime}$ and 0.4$^{\prime\prime}$ were used for the \nisp{}-$H$ and \vis{} channels, respectively. These apertures are optimised to provide the highest signal-to-noise on the stellar flux, by maximising flux within the aperture while minimising blending. The derived magnitudes were compared to the input (\besancon) magnitudes, and a blending co-efficient ($b$) defined as the fraction of the total light in the aperture arising from the host star.

\subsection{Noise model}
\label{ModelRedSect}

Photometric uncertainties can be categorised into random (``white'') and time-dependent (``pink'' or ``red'') components. White noise will reduce as the square root of the number of exposures per transit, $\sqrt{n_{\rm e}}$. Other noise sources will also be reduced by repeated observations, but by a factor depending on their repeatability and characteristic frequency (e.g.\ \citealt{PZQ06}). Many can be removed by high-pass frequency filtering (`whitening') of time-series data. For most time-dependent noise sources, we can assume that the photometric uncertainty limiting our ability to detect transiting objects will decrease as the number of observed transits, $\sqrt{n_{\rm t}}$. White and pink noise, reduced appropriately, are here summed in quadrature to give the total photometric uncertainty. Both noise estimates have been chosen to match the assumptions of Paper~I.


White noise ($\sigma_\ast$) is taken simply to be the detector Poisson noise (accounting for stellar blending). A pink noise estimate of $\sigma_{\rm R}$ = 0.001 has been chosen to cover the time-dependent noise sources. We consider $\sigma_{\rm R}$ = 0.0003 and 0.003 mag as limiting cases, though show this has little effect on the number of detections. While the on-sky amplitude of pink noise is not yet known, our worst-case scenario is extremely conservative, as even ground-based facilities have noise levels well below this. Our estimate includes the following, non-exhaustive list.

{\it Image persistence}. This is likely to be the dominant source of pink noise. A ghost image contrast ratio of $\leq$10$^{-4}$ is set by \euclid{} requirement WL.2.1-23 \citep{LAA+11}. Its effect depends on the repeatability of each pointing. Repeated pointing of ghost images to within the same PSF may fatally degrade a small number of lightcurves. Less-accurate pointing will provide occassional (flaggable) bad data in many lightcurves.

{\it Zodiacal light variation}. Variations in background zodiacal light can be limited to $<$1 per cent over $>$10 years \citep{LBH+98}. The error in its removal will be substantially smaller.

{\it Thermal variation and attitude correction}. The spacecraft--Sun angle is limited to between --1$^\circ$ and +31$^\circ$ as measured orthogonal to the optical axis. Variation in this angle may cause detector thermal noise variations over each month-long observing window. These variations should be reproducible and hence removable (cf.\ the ramp in \emph{Spitzer} IRAC photometry, e.g.\ \citealt{CHH+11,NHH+11,BKB+11,BHM+11}).

{\it PSF changes}. The PSF stability required of \euclid{} should mean PSF changes contribute negligbly to the total error.

{\it Non-periodic or quasi-periodic stellar variability}. Cooler dwarfs show quasi-periodic variability on timescales of days to weeks due to starspots, and on shorter timescales due to flares. \citet{CvBB+11} suggest that M dwarf variability over month-long timescales is typically $<$1 milli-magnitude (mmag) for $\sim$68 per cent of M dwarfs and 1--10 mmag for a further $\sim$25 per cent. We mostly detect planets around higher-mass stars, so this effect will be largely limited to objects blended with M dwarfs.


\section{Detection thresholds}
\label{SNSect}

In this Section, we describe the mathematical model behind our signal-to-noise calculations, and show how the signal-to-noise of the amplitude of the transit, secondary eclipse, orbital modulation, ellipsoidal variation and Doppler boosting are calculated. This is intended as a summary only, as the basic physics can be found in textbook works (e.g.\ \citealt{Perryman11}).

\subsection{Transit detection thresholds}
\label{ModelDetectSect}

\subsubsection{Transit signal-to-noise}
\label{ModelDetectTransSect}

We begin with a simple model, defining the transit as a step function. This neglects: (1) the ingress and egress time of the transit, (2) variation due to limb darkening of the star, and (3) emission from the secondary, which we detail in Sections \ref{ModDistGrazeSect} and \ref{ModPlanetSect}.

The photometric depth ($\Delta_{\rm t}$) and length of the transit are given in the usual way (e.g.\ \citealt{Perryman11}, his eq.\ (6.1) and (6.4)), with the depth corrected by the blending factor, $b$. White ($\sigma_\ast$) and pink ($\sigma_{\rm R}$) noise estimates (Section \ref{ModelRedSect}), normalised by the number of transits ($n_{\rm t}$) and exposures per transit ($n_{\rm e}$), are quadrature summed to give the total noise in the folded transit lightcurve ($\sigma_{\rm t}$), giving a signal-to-noise of:
\begin{equation}
	s_{\rm t} = \Delta_{\rm t} / \sigma_{\rm t} = \left. {\Delta_{\rm t}} \middle/ {\sqrt{ \frac{\sigma_{\ast}^2}{n_{\rm t} n_{\rm e}} + \frac{\sigma_{\rm R}^2}{n_{\rm t}} }} \right. .
\label{sigmatEq}
\end{equation}
We assert that a transit is detectable if the systemic flux in transit is $>$10$\sigma$ lower than the out-of-transit flux in the phase-folded lightcurve, hence if $s_{\rm t} > 10$. This considerably exceeds the minimum value of 6$\sigma$ found by \citet{KZM02}. For comparison, \citet{BKB+08} cite an 84 per cent detection rate for 8$\sigma$ transits in synthetic \kepler{} folded light curves. We therefore believe this limit is suitably conservative and would allow an accurate timing solution to be found for the orbit.

\subsubsection{Grazing transits and limb darkening}
\label{ModDistGrazeSect}

A multiplicative correction can be applied to this simple model to take into account grazing transits and limb darkening. It can be shown from basic geometry (e.g.\ \citealt{Perryman11}; his section 6.4.2) that this is given by:
\begin{equation}
f_{\rm gr} = 1 + \left( \int_{-1-r_{\rm p}/r_\ast}^{1+r_{\rm p}/r_\ast} f_0(x) \, D(x) \, \frac{r_\ast}{r_{\rm p}} \ {\rm d}x \middle/ \sqrt{1-y^2} \right) ,
\label{corfac}
\end{equation}
where $r_\ast$ is the stellar radius, $x$ is the orbital distance from mid-transit and $y$ is the transit impact parameter (both in units of $r_\ast$), $f_0$ is the fraction of the secondary eclipsing the primary star (mathematically given by the intersection of two circles) and $D(x)$ is the mean limb-darkening co-efficient for the eclipsed part. The signal-to-noise of the transit will scale as $\sqrt{f_{\rm gr}}$.

The nature of $f_{\rm gr}$ means it is not a trivial correction to calculate analytically (see, e.g., \citealt{MA02}). We thus incorporate this into our model by numerical integration, and typically find it causes a $<$1 per cent variation on $s_{\rm t}$, thus the exact formulation of the limb-darkening law does not matter significantly. As we show later, most of our detections will be around solar-type stars, thus we approximate $D(x)$ by \citep{Russell1912}:
\begin{equation}
D(x) \approx \frac{1 - w (1 - \cos B)}{n_{\rm D}}
\label{LimbEq}
\end{equation}
where $B = \sqrt{1-(x^2+y^2)}$ and $w \approx 0.41$ for \vis{} and $w \approx 0.20$ for \nisp{}-$H$. These values for $w$ should be accurate to within $\pm 0.1$ for the majority of our stars \citep{vanHamme93}. The factor $n_{\rm D}$ is a normalisation constant such that the average of $D(x)$ over the stellar surface is unity.

The few percent of transiting secondaries with impact factors between $r_\ast$ and $r_\ast + r_{\rm p}$ produce grazing transits of their hosts. These do not produce characteristic flat-bottommed light curves and are therefore much more likely to be mistaken for false positives (blended binary stars) than other transits, adding some additional uncertainty to our detectable planet number. 

\subsection{Planetary reflection and emission}
\label{ModelEmitSect}

Outside of transit, lightcurves are modulated by the difference in emission and reflection from companions' day and night sides (e.g.\ \citealt{JD03,BKJ+09,GVG+10,DSM+11}; see also, e.g., \citealt{Perryman11}, his section 6.4.7), the effect of which can also be seen at secondary eclipse. We model the secondary as a grey, diffusively-scattering (i.e.\ partially-reflecting, but otherwise Lambertian) disc with albedo $A$. The flux from the secondary ($f_{\rm p}$) at a given point in its orbit will be the sum of reflected starlight and emission from its day and night sides, which we model to have uniform but potentially different temperatures using:
\begin{equation}
\label{fpEq}
	\frac{f_{\rm p}}{f_\ast} = 4 A \frac{\pi r_{\rm p}^2}{4 \pi r_{\rm o}^2} \, \frac{\zeta}{\pi} + f_{\rm d}  \frac{1+\cos\theta}{2} + f_{\rm n} \frac{1+\cos(\theta+\pi)}{2} ,
\end{equation}
where $f_\ast$ is the flux from the star, $f_{\rm d}$ and $f_{\rm n}$ are the flux from the secondary's day- and night-side hemispheres, $r_{\rm o}$ is the orbital radius, $\theta$ is the orbital phase from eclipse, $\zeta = \sin\theta+(\pi-\theta)\cos\theta$, and where the initial factor of four in Eq.\ (\ref{fpEq}) arises from the normalisation constant for a Lambertian surface.

\subsubsection{Reflection from the secondary}
\label{ModelDetectDayNightSect}

The reflection term from the secondary (i.e.\ conservatively assuming no emission term), removes the right-hand two terms from Eq.\ (\ref{fpEq}). At secondary eclipse, $\theta \approx \pi$, further reducing it and meaning the eclipse depth can be modelled simply by:
\begin{equation}
	\Delta_{\rm e} \approx \frac{f_{\rm p}}{f_\ast} b \approx A \frac{r_{\rm p}^2}{r_{\rm o}^2} b .
\label{DeltaEEq}
\end{equation}
The signal-to-noise ratio of the eclipse, $s_{\rm e}$ can then be calculated by subsituting $\Delta_{\rm t}$ for $\Delta_{\rm e}$ in Eq.\ (\ref{sigmatEq}). We assume that a timing solution is well-determined by the transit measurement, thus we can define secondary eclipses with $s_{\rm e} >$ 5 to be detections.

For the out-of-eclipse modulation of the lightcuve, we divide the lightcurve between the end of transit (defined as $\theta = \alpha$; Figure \ref{SchemFig}) and eclipse ($\theta = \pi - \alpha$) into two equal portions, corresponding the secondary's gibbous and crescent phases. The flux reflected by the secondary will change between these two periods by (from Eq.\ (\ref{fpEq})):
\begin{equation}
\label{DeltarEq}
	\Delta_{\rm r} \approx \Delta_{\rm e} \frac{\delta f_{\rm cresent} - \delta f_{\rm gibbous}}{\delta f_{\rm crescent}+\delta f_{\rm gibbous}} ,
\end{equation}
where $\delta f_{\rm crescent} = \int_{\alpha}^{\pi/2} \zeta {\rm d} \theta$ and $\delta f_{\rm gibbous} =  \int_{\pi/2}^{\pi-\alpha} \zeta {\rm d} \theta$. The approximation arises because $\alpha \neq 0$. Observed transiting planets have $r_{\rm o} > 3 r_\ast$, meaning we can expect that $\alpha$ is always small ($\sin\alpha < 1/6$), hence the ratio in Eq.\ (\ref{DeltarEq}) should always be between 0.753 and 0.785. For convenience, we assume $\Delta_{\rm r} \approx 0.76 \Delta_{\rm e}$.

The signal-to-noise ratio can be computed using Eq.\ (\ref{sigmatEq}), replacing $\Delta_{\rm t}$ by $\Delta_{\rm r}$ and $n_{\rm e}$ by $n_{\rm e}^\prime = 2 (p_{\rm o} - p_{\rm t}) / 4 t_{\rm c}$, where the initial factor of two accounts for both the waxing and waning phases, $p_{\rm o}$ and $p_{\rm t}$ are the orbital period and transit length and $t_{\rm c}$ is the observing cadence. A final reduction by a factor of $\sqrt{2}$ comes from the quadrature-added errors in $\delta f_{\rm crescent}$ and $\delta f_{\rm gibbous}$. Again, we require $s_{\rm r} > 5$ for a detection.

\subsubsection{Emission from the secondary}
\label{ModPlanetSect}

The composition of the secondary companion alters our sensitivity to its emitted light. Without \emph{a priori} knowledge of the objects' composition, we cannot accurately model planetary emission, hence must accept a significant, presently unquantifiable uncertainty in our ability to detect emission from companions.

The efficiency of atmospheric heat transport ($\epsilon$) determines the equilibrium temperature of the secondary's day and night sides. We model three scenarios: (1) inefficient heat transport, $\epsilon = 0$, whereby the secondary emits only from the day side, hence $f_{\rm n} \approx 0$ (cf.\ CoRoT-7b; \citealt{NSW+10}); (2) efficient heat transport, $\epsilon = 1$, such that the day- and night-side temperatures are equal ($T_{\rm d} = T_{\rm n}$; cf.\ Venus); (3) $\epsilon = 0.5$, such that the $T_{\rm n}/T_{\rm d} = 5/6$ (from $L \propto T^4$), close to the situation on HD 189733b \citep{KCA+07}.

Idealising the star and planet as blackbodies with equilibrium temperatures $T_\ast$ and $T_{\rm p}$, their contrast ratio becomes:
\begin{equation}
R_{\rm c} = \frac{{\rm B}(\lambda,T_\ast)}{{\rm B}(\lambda,T_{\rm p})} \, \frac{r_\ast^2}{r_{\rm p}^2}  = \frac{\exp\left(\frac{hc}{\lambda k_{\rm B} T_{\rm p}}\right) - 1}{\exp\left(\frac{hc}{\lambda k_{\rm B} T_\ast}\right) -1 } \, \frac{r_\ast^2}{r_{\rm p}^2} \equiv \frac{\Lambda(T_\ast)}{\Lambda(T_{\rm p})} \frac{r_\ast^2}{r_{\rm p}^2},
\label{ContrastEq}
\end{equation}
where $\lambda$ ($\approx 880$ nm for \vis{}), $c$ and $k_{\rm B}$ have their usual meanings. For a heat transport efficiency of $\epsilon$, and given $L \propto T^4$, $T_{\rm d}$ and $T_{\rm n}$ are described by:
\begin{equation}
	T_{\rm n} = \epsilon^{1/4} T_{\rm eq} \quad {\rm and} \quad T_{\rm d} = (2 - \epsilon)^{1/4} T_{\rm eq} ,
\label{TdEq}
\end{equation}
where the equilibrium temperature of the secondary is given by:
\begin{equation}
	T_{\rm eq} = T_\ast \left(1-A\right)^{1/4} \left(\frac{r_{\rm o}}{r_\ast}\right)^{-1/2} ,
\label{TeqEq}
\end{equation}
where $A$ is the Bond albedo.

A more-complete model of the fractional depth of secondary eclipse (Eq.\ (\ref{DeltaEEq})) can then be computed as:
\begin{equation}
	\Delta_{\rm e} = \frac{f_{\rm p}}{f_\ast} b = \left( 4 A \frac{\pi r_{\rm p}^2}{4 \pi r_{\rm o}^2} + \frac{\Lambda(T_\ast)}{\Lambda(T_{\rm d})} \, \frac{r_{\rm p}^2}{r_\ast^2} \right) b .
\label{DeltaEEq2}
\end{equation}
The orbital modulation of the lightcurve between cresent and gibbous phases becomes (by modifying our existing approximation) :
\begin{equation}
	\Delta_{\rm r} \approx 0.76 \left( 
		4 A \frac{\pi r_{\rm p}^2}{4 \pi r_{\rm o}^2} +
		R_{\rm dn} R_{\rm ps} \frac{r_{\rm p}^2}{r_\ast^2} 
		\right) b ,
\label{DeltaREq2}
\end{equation}
assuming the secondary's radiation follows a sinusoidal variation between the sub-solar and anti-solar points (to match the Lambertian reflectance). $R_{\rm dn}$ is the difference between the observed day- and night-side flux as a fraction of the secondary's observed phase-average flux, given by:
\begin{equation}
R_{\rm dn} = \frac{\Lambda(T_{\rm d}) - \Lambda(T_{\rm n})}{\Lambda(T_{\rm d}) + \Lambda(T_{\rm n})} ,
\end{equation}
whilst $R_{\rm ps}$ is the ratio of the secondary's phase-averaged flux to the stellar flux, given by:
\begin{equation}
R_{\rm ps} = \frac{ \frac{1}{2} \left( \Lambda(T_{\rm d}) + \Lambda(T_{\rm n})	\right)	}{ \Lambda(T_\ast)} .
\end{equation}
The factor $1/2$ in the above equation comes from the fact we only see one hemisphere (of both bodies) at a given time.

\subsection{Gravitational effects of the secondary}
\label{ModGravitySect}

Tidal distortion of the primary star modulates the stellar lightcurve at twice the orbital frequency. Though weak, it can yield the secondary's mass directly from the lightcurve, confirming an object's planetary nature (e.g.\ \citealt{vKRb+10}). For a circular orbit, aligned to the stellar rotation axis, the projected surface area of the star changes by (\citealt{KM09}, their eq.\ (3.1.53)):
\begin{equation}
E \approx \frac{
		1 + \frac{1}{6} \left(1+7\frac{m_{\rm p}}{m_\ast}\right) \left(\frac{r_\ast}{r_{\rm o}}\right)^3
	}{
		1 + \frac{1}{6} \left(1-2\frac{m_{\rm p}}{m_\ast}\right) \left(\frac{r_\ast}{r_{\rm o}}\right)^3
	} ,
\label{EllipseEq}
\end{equation}
where $m_\ast$ is the stellar mass. The change in stellar flux will be of similar magnitude to $E$. However, a finite observing time is required. Comparing the flux near quadrature ($\pi/4 < \theta < 3\pi/4$) to the flux near eclipse ($3\pi/4 < \theta < 5\pi/4$), yields a change of:
\begin{equation}
\Delta_{\rm E} = E\big(1 - 2(1 - 2\sqrt{2}/\pi)\big) .
\end{equation}
This represents an upper limit, as the transit and eclipse signals will dominate during their respective periods. We take a slightly more conservative estimate, basing our signal-to-noise calculation on $\Delta_{\rm E} = 0.7 E$.

The host star's light is also Doppler boosted as it orbits the system barycentre, leading to a sinusoidal variation on the orbital period, a quarter revolution out of phase with the flux reflected or emitted by the secondary's day side. Its amplitude depends on both the stellar reflex velocity, $v_{\rm r}$, inclination of the secondary's orbit to our line of sight, $i$, and the spectral index of the star in that band, $\alpha_{\rm i} = {\rm d\ ln}\, F_\nu / {\rm d\ ln}\  \nu$ \citep{BK79}:
\begin{equation}
\Delta_{\rm D} = \left( \frac{\sqrt{1- v_{\rm r}^2/c^2}}{1- (v_{\rm r}/c) \cos i} \right) ^{3-\alpha_{\rm i}} .
\label{DopplerEq}
\end{equation}
Around the $H$-band H$^{-}$ opacity peak, $\alpha \approx 1.0$. For the \vis{} detections, $\alpha$ is typically between --0.1 and --0.7. We assume $\alpha$ = --0.4.


\section{Simulating the companion population}
\label{ExoSect}

To simulate the observable secondary population, we begin with an underlying model that makes the fewest assumptions and approximations about the underlying population as possible. We then explore the limits that current theory and observations can place on a more sophisticated model, and use these to assign an error budget.

\subsection{A simple binary model}
\label{ModelPlanSect}

\subsubsection{Generating orbits}
\label{ModelPlanOrbitSect}

Orbits are randomly selected from a logarithmic distribution in radius. This is bounded by $r_{\rm o} \geq 3 r_\ast$, which approximately corresponds to the observed closest-orbiting planets (e.g.\ \citealt{HCCL+09,CCGS+10}) and a requirement that at least three transits are observed (i.e.\ $p_{\rm o} \leq 493$ days). A random orbital inclination ($i$) and random orbital phase ($\psi$) with relation to the coverage period are assigned. We assume circular orbits in all cases. Equilibrium temperatures of the secondaries are calculated assuming a geometric albedo of $A = 0$.

\subsubsection{Generating companion radii}
\label{ModelPlanRadiiSect}

Secondary companion masses are randomly selected from a logarithmic distribution between 0.00316 and 316 Jupiter masses (M$_{\rm J}$), extending from Earth mass to well into the stellar regime.

\begin{figure}
\centerline{\includegraphics[height=0.50\textwidth,angle=-90]{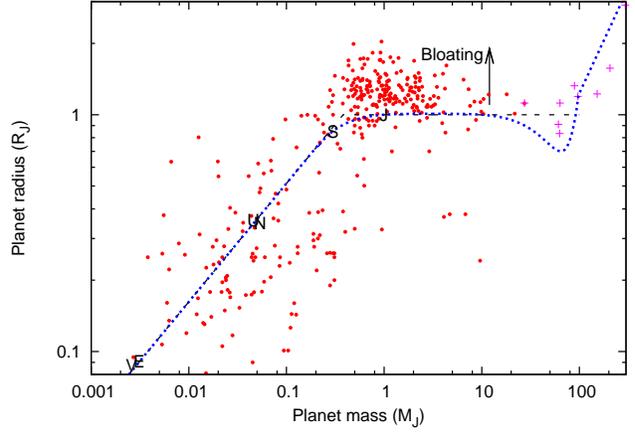}}
\caption{Adopted first-order and second-order mass--radius relations (black dashed and blue dotted lines, respectively). The lower boundary of these lines denotes the assumed relation for objects with equilibrium temperatures below 1000 K, with bloating at higher temperatures increasing object radius. Points show known objects (see text). Letters denote Solar System objects.}
\label{MRFig}
\end{figure}

The mass--radius ($m_{\rm p}$--$r_{\rm p}$) relation depends strongly on the secondary's internal structure and composition. Nevertheless, observations suggest that the relation approximates a power-law in the sub-Jupiter-mass regime, while super-Jupiter-mass planets and brown dwarfs have roughly constant radii. Mass and radius then become roughly proportional at higher masses \citep{CBL+09}. Figure \ref{MRFig} shows this relation for transiting exoplanets (\protect\citealt{SDLS+11}), brown dwarfs\footnote{WASP-30b, \citep{ACCH+11}; CoRot-15b, \citep{BDG+11}; LHS 6343C, \citep{JAG+11}; Kelt-1b \citep{SBP+12}; and OGLE-TR-122b and -123b \citep{PMB+05,PMB+06}.} and low-mass stars\footnote{GJ551 (Proxima Centauri), GJ699 and GJ191 \protect\citep{SKFQ03}.}.

\begin{figure}
\centerline{\includegraphics[height=0.50\textwidth,angle=-90]{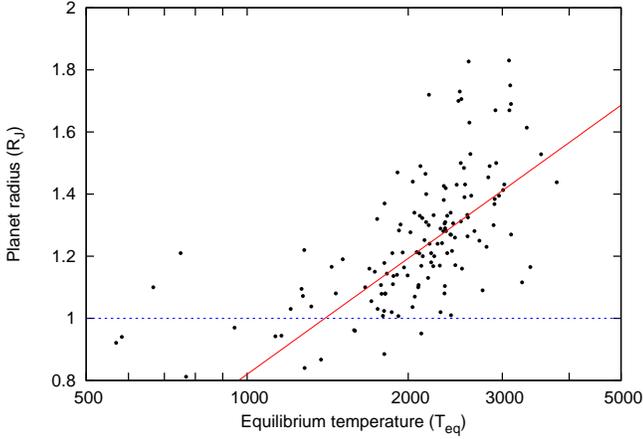}}
\caption{Equilibrium temperature of known exoplanets versus planetary radius for objects over 0.6 M$_{\rm Jup}$. A fit is shown to planets with $T_{\rm eq} > 1000$ K. The dashed line marks 1 R$_{\rm Jup}$, which we use as a lower limit.}
\label{BloatFig}
\end{figure}

It is recognised, however, that some planets have much larger radii because internal thermal pressure acts against gravitational contraction. Strongly irradiated planets, typically in short orbits around young stars, are particularly bloated (e.g.\ \citealt{LCBL10,FDD+11}). \citet{FDD+11} identifies bloating as being effective for $T_{\rm eq} \gtrsim 1000$ K. Using an updated version of the same data source\footnote{http://exoplanets.org/}, we model the radius for known planets with $T_{\rm eq} > 1000$ K and $m_{\rm p} > 0.6$ M$_{\rm Jup}$ planets. A simple linear regression (Figure \ref{BloatFig}) gives:
\begin{equation}
r_{\rm p} = (1.239 \pm 0.154) \log(T_{\rm eq}) - (2.897 \pm 0.514) .
\end{equation}

We account for bloating using a function which mimics the isochrones of \citet{CBL+09} in the limit of an old ($\sim$5--13 Gyr) population, as is suitable for the Galactic Bulge:
\begin{equation}
	\frac{ r_{\rm p} }{ {\rm R}_{\rm Jup} } = \left\{ 
		\begin{array}{rl}
\vspace{2mm}
B \sqrt{318 \frac{m}{{\rm M}_{\rm Jup}}}  / 11							&\mbox{ if $m <$ 0.22 M$_{\rm Jup}$} \\
\vspace{2mm}
\displaystyle
B \Biggl( 1.009 - \frac{(\log_{10}(m_{\rm p}/{\rm M}_{\rm Jup})-0.5)^4}{7} \cdots
\!\!\!\!\!\!\!\!\!\!\!\!\!\!\!\!\!\!\!\!\!\!\!\!\!\!\!\!\!\!\!\!\!\!\!\!\!\!\!\!\!\!\!\!\!\!\! \\
\displaystyle
\cdots + \left(\frac{m_{\rm p}/{\rm M}_{\rm Jup}}{100}\right)^5 \Biggr)		&\mbox{ if $0.22 \!\leq\! m \!<\! 100$ M$_{\rm Jup}$ } \\
\vspace{2mm}
B (m_{\rm p} / {\rm M_\odot})^{8/9}\frac{\rm R_{\odot}}{\rm R_{\rm Jup}}	&\mbox{ if $m \geq$ 100 M$_{\rm Jup}$ ,} \\

		\end{array} \right.
\label{MRRelEq2}
\end{equation}
where $B$ is the bloating factor, given by:
\begin{equation}
	B = \left\{
		\begin{array}{rl}
			\vspace{2mm}
			1				&\mbox{ if $T_{\rm eq} <$ 1398 K} \\
			\vspace{2mm}
			1.239 \log T_{\rm eq} - 2.897	&\mbox{ if $T_{\rm eq} \geq$ 1398 K} \\
		\end{array} \right. .
\label{BloatEq}
\end{equation}

For each star, a sufficient number of random companions were modelled to reach a number density of planets per star that matches observations. For our initial, simple model, we assume 1/30 planets exist per factor of ten in planetary mass and orbital period. The change from 1/3 planets per dex$^2$ used in Paper~I reflects the difference in observed numbers of transiting versus microlensed planets, and is examined more closely in the next Section.

\subsection{Factors modifying the intrinsic distribution of companions}
\label{ModDistSect}

Several intrinsic factors may change the distribution of our modelled systems. We describe major contributors below and our sensitivities to them. Our adopted values in each case are described at the end of this sub-section.

\subsubsection{Secondary mass and period distributions}
\label{ModDistDistSect}

We can formulate a double power law describing the variation of planet frequency ($f$) with secondary mass (or radius) and orbital period as:
\begin{equation}
\frac{\partial^2 f}{\partial \ln m \partial \ln p_{\rm o}} = C_{\rm m,p} \left( \frac{m_{\rm p}}{m_{\rm ref}} \right)^{\alpha_{\rm m}} \left( \frac{p_{\rm o}}{p_{\rm ref}} \right)^\beta \quad {\rm , or}
\label{PlDensityEq}
\end{equation}
\begin{equation}
\frac{\partial^2 f}{\partial \ln r_{\rm p} \partial \ln p_{\rm o}} = C_{\rm r,p} \left( \frac{r_{\rm p}}{r_{\rm ref}} \right)^{\alpha_{\rm r}} \left( \frac{p_{\rm o}}{p_{\rm ref}} \right)^\beta ,
\label{PlDensityEq2}
\end{equation}
for arbitrary reference values $m_{\rm ref}$, $r_{\rm ref}$ and $p_{\rm ref}$. $C_{\rm m,p}$ must be defined over a given range of masses and orbital radii, given the reference mass and orbital period used. For our simple model, which adopts a secondary number density in which $\alpha_{\rm m} = \beta = 0$, we simply set $C_{\rm m,p} = 1/30$ planets dex$^{-2}$ star$^{-1}$.

Observationally, the exponents $\alpha_{\rm m}$ and $\beta$ are poorly determined, but there is a general consensus that low-mass, wider-orbit planets are more common. \citet{CBM+08} adopts values of $\alpha_{\rm m} = -0.31 \pm 0.2$ and $\beta = 0.26 \pm 0.1$ in the regime $0.3 < m_{\rm p} < 10$ M$_{\rm Jup}$ and $2 < p_{\rm o} < 2000$ days, with the constant $C_{\rm m,p}$ set such that 10.5 per cent of stars have planets in this range (equating to $C_{\rm m,p}$ = 0.023 planets dex$^{-2}$ star$^{-1}$ for $\alpha_{\rm m} = \beta = 0$).

\citet{Youdin11} derives a number of values of $\alpha_{\rm r}$ and $\beta$ from \kepler{} planetary candidates, showing that $\alpha_{\rm m}$ and $\beta$ vary by $\pm \sim \! 1$ for small and short-period planets. This may mean tidal orbital decay is important (we discuss this in Section \ref{ModDistTideSect}). In any case, for large planets and sub-stellar objects, $\alpha_{\rm r}$ is a poor parameterisation, as planetary radius becomes very weakly dependent on mass for objects above Saturn mass (Figure \ref{MRFig}). Adopting Youdin's parameterisations does not yield realistic number estimates for our input model, as the radius range we are sensitive to is considerably different to the \kepler{} sample. For our analysis, we therefore adopt Eq.\ (\ref{PlDensityEq}).

\subsubsection{Scaling of companion frequency to different host masses and metallicities}
\label{ModDistMassSect}

Evidence is emerging that the incidence of large planets around low-mass stars is less than around high-mass stars \citep{JBM+07,ES10}, particularly as one enters the M-dwarf regime \citep{ECK+06,BJM+06}. This is perhaps unsurprising as one would expect less-massive stars have less-massive planet-forming discs. While a limited number of microlensing results may contradict this idea (e.g.\ \citealt{DBG+09,KSC+13,SCT+13}), this is not universally true of microlensing results \citep{CG14}.

It is also known that the occurrence of massive planets is highly metallicity-dependent, with a much stronger prevalence of such planets around super-solar metallicity host stars \citep{FV05,JBM+07,US07,GL07}, with few (if any) exoplanet hosts with [Fe/H] $<$ --0.5 \citep{STL+09,SDLS+11,MSS+12}. This has specifically been shown in the Galactic Bulge \citep{BSA+10}. Conversely, for low-mass planets (below $\sim$30 M$_\oplus$), occurence seems to be largely metallicity-independent \citep{MML+11,DBF+12,BLJ+12}. Consequently, the absolute scaling of companion frequency with metallicity is still very poorly determined and depends on the parameters of individual surveys.

We can modify Eq. (\ref{PlDensityEq}) to include a mass and metallicity ($Z_\ast$) scaling as follows: 
\begin{multline}
\frac{\partial^4 f}{\partial \ln m_{\rm p} \partial \ln p_{\rm o} \partial \ln m_\ast \partial \ln Z_\ast} = \\
C_{\rm m,p,m\ast,Z} \left( \frac{m_{\rm p}}{m_{\rm ref}} \right)^{\alpha_{\rm m}} \left( \frac{p_{\rm o}}{p_{\rm ref}} \right)^\beta \left( \frac{m_\ast}{M_\odot} \right)^\gamma \left( \frac{Z_\ast}{Z_\odot} \right)^\delta .
\label{PlDensityEq4}
\end{multline} 
While $\gamma$ is poorly defined, the general consensus of the papers listed above is that $\gamma \sim 1$. The value of $\delta$ is known a little more precisely: \citet{GL07} find that $\delta = 2.22 \pm 0.39$ for stars at $d < 25$ pc and $\delta = 3.00 \pm 0.46$ for stars at $25 < d < 50$ pc.

\begin{figure}
\centerline{\includegraphics[height=0.50\textwidth,angle=-90]{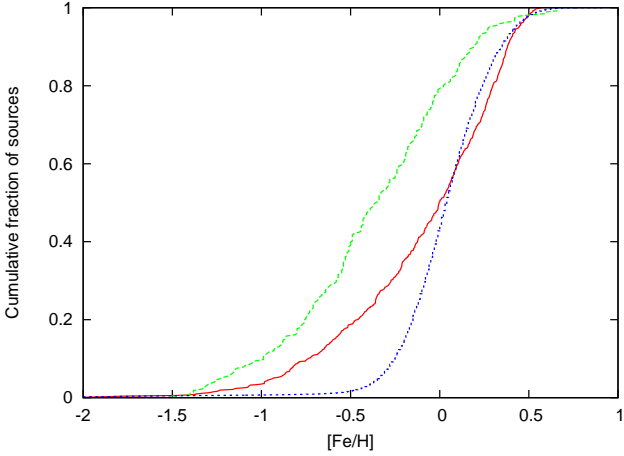}}
\caption{Metallicity distribution of Bulge stars in the \besancon\ models (blue, short-dashed line) versus observations from SDSS/APOGEE DR10, limited to $-4.5^\circ < b < 4.5^\circ$ (red, solid line) and off-axis Bulge fields at higher galactic latitudes from \citet{JRK+13} (green, dashed line).}
\label{ZFig}
\end{figure}

At $<$[Fe/H]$>$ = 0.03 dex, the mean metallicity of the stars simulated by the \besancon\ models differs little from solar\footnote{We use [Fe/H] as a proxy for metallicity ($Z$) in this work. Metal-poor stars typically have enhanced [$\alpha$/Fe], but at levels that do not grossly affect our results. Elemental abundances for solar-metallicity stars in the Bulge are typically similar to solar values \citep{JRK+12,JRK+13,JRK+14}.}. There is, however, a significant dispersion of metallicities, including a small population of thick disc and Halo stars at low metallicity (see Figure \ref{ZFig}).

The distribution of metallicities in the \besancon\ models is a little smaller than that observed in the Bulge\footnote{The newest version of the \besancon\ model includes two Bulge components}, which show a larger metal-poor tail. While no large-scale metallicity analyses have been performed in our survey area, the Sloan Digital Sky Survey Apache Point Observatory Galactic Evolution Experiment Data Release 10 (SDSS/APOGEE DR10) includes stars from nearby Bulge fields. These are mostly expected to be Bulge giants, and show an additional metal-poor component compared to the \besancon\ models, representing about 30 per cent of stars. A specific survey of Bulge giants has been published by \citet{JRK+13}. While this was performed at higher Galactic latitude where stars are expected to be more-metal-poor, it broadly repeats the range of metallicities present in SDSS/APOGEE DR10.

Despite these differences between the metallicity distribution of the \besancon\ models and recent surveys, there should be surprisingly little effect on planet numbers as the metal-rich stars (around which most planets can be expected) are well modelled by the \besancon\ models. Numerical modelling of the SDSS results suggests the total number may be increased by $\sim$1--6 per cent, depending on the value of $\delta$. For a conservative approach, we retain the planet numbers obtained from the \besancon\ model without further alteration.

\subsubsection{Companion destruction via orbital migration}
\label{ModDistTideSect}

Tidal decay can lead to the destruction of companions but, more significantly, it can substantially alter the frequency of companions in very short orbits. This may partially explain the differences \citet{Youdin11} finds in short- and long-period \kepler{} planetary candidates. Orbits of close companions can tidally decay into their host stars over timescales of \citep{BO09}:
\begin{eqnarray}
\tau_{\rm a} &\approx& 12 \, {\rm Myr} 
	\left( \frac{Q_1}{10^6}                       \right)                  
	\left( \frac{m_\ast}{{\rm M}_\odot}           \right)^{8/3}  
	\left( \frac{m_{\rm p}}{{\rm M}_{\rm Jup}}    \right)^{-1}           
	\left( \frac{r_\ast}{{\rm R}_\odot}           \right)^{-5}           \cdot \nonumber\\
&&	\left( \frac{p_{\rm o}}{{\rm 1\ day}}         \right)^{13/3} 
	\left( 1 - \frac{p_{\rm o}}{p_\ast}           \right)^{-1}           
\label{TInspEq}
\end{eqnarray}
where $Q_1$ is the tidal quality factor\footnote{The factor $Q^\prime_1$ in \protect\citet{BO09} reduces to $Q_1$ for an idealised homogeneous fluid body.} and $p_\ast$ is the rotation period of the star. We assume this follows eq.\ (7) of \citet{CC07}, namely:
\begin{equation}
p_\ast = 25.6 \ {\rm days} \ \left(\frac{\tau_\ast}{\rm Gyr}\right)^{0.45} \left(1.4 - \frac{m_\ast}{{\rm M}_\odot}\right),
\label{ProtEq}
\end{equation}
where $\tau_\ast$ is the age of the star in Gyr. This relation is valid for stars below the Kraft break ($m_\ast \lesssim 1.3$ M$_\odot$; \citealt{Kraft67}), encompassing most of our detected host stars. We take $\tau_\ast$ in this case to be half the age of the star at the time of observation. In practice, the value of $p_\ast$ does not much affect the inspiral timescale provided $p_{\rm o} / p_\ast \ll 1$. For most cases, evolution of $p_\ast$ is not affected by the evolution of the companion's orbit, as magnetic braking dominates over transfer of orbital to rotational angular momentum \citep{BO09}. As $\dot{p}_{\rm o}/p_{\rm o} \propto p_{\rm o}^{-13/3}$, there is a rapid change from a quasi-stable state to planetary destruction. We can therefore approximate inspiral by simply removing secondaries where the systemic age is greater than the inspiral time.

Short-period planets may still be observable if they have been recently scattered into their present orbits and not had time to decay (e.g.\ via three-body gravitational interaction or the Lidov--Kozai mechanism; \citealt{Kozai62,Lidov62}). This is a possible explanation for the large number of retrograde, elliptical and highly-inclined orbits of hot Jupiters we see (see \citealt{NFL+10}; e.g.\ WASP-33, \citealt{CCGS+10}). Our estimates of the fraction of decaying companions are thus likely to be an over-estimate. Orbital decay may also be prevented in rare cases of tidal locking ($p_\ast = p_{\rm o}$) or if companions are trapped in orbits which are resonant with a harmonic of the host star's natural frequency (again, cf.\ WASP-33; \citealt{HMRN10}).

$Q_1$ is poorly determined and is likely to depend strongly on spectral type. It is generally thought to lie in the range $10^5$ to $10^{10}$, with values of $10^7$ to $10^{8.5}$ suggested by recent works (\citealt{PJST12}; \citealt{Taylor14}) . A number of known exoplanets appear to give lower limits to $Q_1$. For instance, WASP-3b, WASP-10b, WASP-12b and WASP-14b require $Q_1 \gtrsim 1 \times 10^7$, $3 \times 10^5$, $5 \times 10^8$ and $2 \times 10^7$, respectively, but of these only WASP-3b is not in an eccentric or inclined orbit, making it uncertain whether the planets have always occupied these orbits \citep{PSCC+08,CGS+09,HCCL+09,JPCC+09}. Estimates of $Q_1$ from transit timing data suggest a value of $\sim$10$^7$ is most likely \citep{LDG11,HACC+11}, but direct determination of $Q_1$ for a variety of stellar types has so far been lacking due to the long time bases needed.

We use the observed ``3-day pile-up'' of known exoplanets to model a value for $Q_1$. That is, we shape our modelled companion distribution by varying $Q_1$ to reproduce the observed peak in companion numbers that falls between 2.5 and 4.0 days \citep{WMR07}. We find this is well-matched by values of $\log Q_1^\prime$ lying between 6.7 and 8.0. While this may not represent the true uncertainty on $Q_1$, its correlation with $\beta$ means that any additional uncertainty is incorporated within the \kepler{} results used to define the uncertainty on $\beta$. We thus adopt $\log Q_1 = 7.0$ and 8.0 as our limiting cases and take $\log Q_1$ = 7.5 as our most-likely value.

\subsubsection{Adopted values}
\label{ModDistAdoptSect}

For our physical estimates, we adopt a scaling of planet frequency derived from Eq.\ (\ref{PlDensityEq2}), incorporating tidal decay using Eq.\ (\ref{TInspEq}). The constants in these equations are set as described in Table \ref{ParamTable}, on the basis of the literature described in this sub-section. We compute a best-estimate model using the central values for each of the these parameters. We account for the uncertainties by computing additional Monte-Carlo models. For each of these additional models, each parameter is randomly chosen from a Gaussian distribution with the central value and characteristic width shown, truncated at the limiting values shown. We also adopt a random red noise level, taken from a log-normal distribution between 300 and 3000 parts per million (1000 ppm for our best-estimate model).

\begin{center}
\begin{table}
\caption{Adopted parameters for our best-fit and limiting models}
\label{ParamTable}
\begin{tabular}{lrrrr}
    \hline \hline
Parameter	& Best-estimate	& Gaussian	& Lower 	& Higher \\
\ 		& value		& $\sigma$	& cutoff 	& cutoff \\
    \hline
$\alpha_{\rm m}$	& --0.31 & 0.20 & --0.5 & 0.0 \\
$\beta$			&   0.26 & 0.10 &   0.0 & 0.5 \\
$\gamma$		&   1.0  & 0.5  &   0.0 & 1.5 \\
$\delta$		&   2.6  & 0.3  &   0.0 & 3.5 \\
$\log(Q_1)$		&   7.5  & 1.0  &   7.0 & 8.5 \\
    \hline
\end{tabular}
\end{table}
\end{center}

This allows us to adopt a range in detected planet numbers whereby 68 per cent of models have a number of detected planets within that range.

\subsection{Resolving stellar blends}
\label{BlendsSect}

We reject stars which are significantly blended with others, though the point at which a star is `significantly blended' is an arbitrary choice. Our nominal scenario rejects stars where $<$75 per cent of the light within the photometric aperture comes from the host star ($b < 0.75$) for the \vis{} data. We choose $b < 0.5$ for \nisp{}-$H$, on the basis that these blends will be discernible in the \vis{} photometry. This is a conservative estimate, and many of these may be recoverable, either through difference imaging or PSF-fitted photometry. Nevertheless, most potential detections qualify as unblended. We also compute detections for $b > 0.5$ and 0.25 (\vis{} and \nisp{}-$H$, respectively) as our `best-case' scenario.

A key determinant of our ability to resolve these stellar blends is the distance from the blended host star to the blending star in question. In our simulated data, 0.4 per cent of stars have a nearest neighbour closer than 0.3$^{\prime\prime}$ (one $NISP$ pixel or three \vis{} pixels) and only 2.0 per cent of stars have a neighbour closer than 0.6$^{\prime\prime}$. Many of these blends are with fainter stars. The full-width half-maximum (FWHM) of the \vis{} PSF is 0.4$^{\prime\prime}$, meaning that the majority of blended light comes from the PSF wings of much brighter stars at comparatively large angular distances.

\subsection{False positive rejection}
\label{FalsePosSect}

We have adopted conservative parameters throughout our analysis. However, many transit-like events will also be caused by false positives: mostly blended eclipsing binaries (BEBs), star spots and seismic activity and pulsation of `isolated' stars.

Modelling false positives is inherently complex and requires better knowledge of many unknowns of both the survey and observed stellar population. Indeed, publication of planets from similar surveys are typically released initially as `objects of interest': lists of candidate planets for future follow-up (e.g.\ \citealt{LWW+07,SCC+07,KCW+08,BRB+12}), where false positives are excluded to the best of the survey's ability, rather than confirmed planets included when false positives have been reasonably ruled out. For these reasons, we do not model the effect of false positives on our dataset. Instead, we detail methods to reject false positives from the data, some of which can be improved upon with additional ground-based follow-up data, and show that few false positives are likely to contaminate the detections of transiting companions.

\subsubsection{Blended eclipsing binaries}
\label{FalseBEBsSect}

BEBs can mimic much-smaller objects transiting isolated stars, as they dilute the depth of the eclipse. Often referred to as background eclipsing binaries, BEBs may well be intrinsically-fainter foreground binaries or hierarchical triple (or higher multiple) systems. A major advantage of a \vis{}~+~\nisp{}-H survey with \euclid{} is that many of these systems will be discernable by a combination of several methods.

{\it Direct resolution:} Like blended transiting systems, BEBs will be directly resolvable from field stars in the majority of cases. We expect this to be the primary method of resolving blended eclipsing binaries at seperations $\gtrsim$0.5$^{\prime\prime}$.

{\it Astrometric shift during eclipse:} The fine resolution of \vis{} and the good global astrometric solution should mean the centre of light of the blended systems will change appreciablely. The 0.1$^{\prime\prime}$ resolution of \vis{} may allow detection of centroid changes down to $\approx$3 milli-arcseconds, equivalent to a 1 per cent dip in total light caused by a BEB which is 0.3$^{\prime\prime}$ from a third, brighter star.

{\it Colour change during eclipse:} This is effective for stars with two components of different \vis{}--$H$ colours and benefits from the dual-colour observations. A blend of two stars of different colours (such as a G star and a similar-mass, M-star binary) will change colour significantly: there will be little change in the \vis{} flux, but significant change in the $H$-band, where the eclipsed M star is relatively much brighter. Its success depends on both the magnitude and colour differences between the BEB, or between the BEB and blended single star. For example, a 1.2 M$_\odot$ F dwarf with a blended 0.7+0.7 M$_\odot$ mutually-eclipsing K dwarf binary at the same distance would exhibit eclipses $\sim$30 per cent deeper in $H$-band than \vis{}.

{\it Ellipsoidal variability:} With higher-mass secondaries, BEBs are more ellipsoidally distorted than isolated transiting systems. In blended systems, the strength of ellipsoidal variability should also be different in $H$-band than \vis{}.

{\it Ingress/egress length:} BEBs will take longer to complete ingress and egress than Jupiter-radius sub-stellar objects.

{\it Stellar density profiling:} If a measure of stellar density can be obtained, comparison with the stellar density obtained directly from the transit lightcurve can be used to identify background eclipsing binaries via the photo-blend or photo-timing technique \citep{Kipping14}. Stellar density is best obtained through asteroseismology. We may have sufficient signal-to-noise to perform asteroseismology to (e.g.) separate giants from main-sequence stars, but we are limited by our frequency coverage. If the satellite telemetry rate could be increased above the presently agreed rate, such that the \vis{} cadence could equal the \nisp{}-$H$ cadence of 1095 seconds, our Nyquist frequency would increase from 139 $\mu$Hz to 457 $\mu$Hz. This would greatly improve our sensitivity towards more-solar-like stars, though we would not be able to perform full asteroseismological analysis of our expected planet hosts. Density can also be derived from a measure of surface gravity: traditionally through ionisation equilibrium of Fe {\sc i} and {\sc ii} equivalent line widths combined with a measure of stellar temperature and/or luminosity (cf.\ \citealt{JP10,MJZ11}), though the stars will likely be too faint. For \exels{}, a powerful tool may be the use of stellar flicker on the timescales of hours \citep{BSBP13,KBS+14}, combined with an extinction-corrected photometric luminosity measurement (e.g.\ \citealt{MvLD+09,MZB12}).

\subsubsection{Stellar pulsations and star spots}
\label{FalsePulsSect}

Small-amplitude pulsations, particularly among giant stars or background blends, can mimic repeated transits in low signal-to-noise data. We do not expect these to be a major contaminant. Firstly, there is a lack of faint background giant stars: only 0.18 per cent of modelled stars have $T_\ast < 5\,500$ K and $m_{\rm bol} < 2$ mag, at which point pulsation in the \vis{} band is expected to be $\ll$10 mmag \citep{MZS+14}. Secondly, as with \kepler{}, the low red noise measure means that the classic U-shaped transit signal should be differentiable from the more-sinusoidal stellar pulsations.

The multi-year observing plan of \exels{} mitigates to some extent against the confusion of star spots with transiting bodies, as starspots will vary in size and drift in longitude during this period. A quantitatively robust method of distinguishing against star spots could come from comparative density measurements (as for the BEBs above), using the photo-spot method of \citet{Kipping14}.

\subsection{Measuring the secondary's mass}
\label{FalseHMRsSect}

Ideally, we would like to separate planet-mass companions from brown dwarfs or low-mass stars by robustly measuring the secondary's mass. Statistically, the comparatively small numbers of brown dwarfs, the so-called ``brown dwarf desert'' \citep{MB98,HAM+00,DDA+08}, will favour planet-mass bodies. Still, a quantitative determination of a companion's status as a planet must lie either in radial velocity variations (which we deal with in Section \ref{FollowSpectSect}), a measure of the body's internal heat generation, detection of ellipsoidal variation or Doppler boosting commensurate with a higher-mass object, or the transit-timing variations caused by  other bodies in the system (cf.\ Kepler-11; \citealt{LFF+11}). These measures will be prohibitively difficult for fainter objects, thus we will not be able to differentiate star--brown-dwarf from star--planet systems, except in the relatively rare case where one of these methods can be successful. The photo-mass approach, a combination of techniques by which the mass can be recovered from the lightcurve \citep{Kipping14}, may be useful in separating the largest-mass companions, though it will be at the limit of this method's sensitivity. More general statistical validation techniques may represent our best chance to directly infer that candidates are truly planets (e.g.\ \citealt{Morton12,DAS+14}).

Our approach prevents us from accurately investigating the number of multiple-planet systems we can expect to find, thus we cannot model the effect of transit-timing variations. We have also neglected internal heat generation in our calculations, thus cannot accurately model our use of this technique. We discuss the use of ellipsoidal variation and Doppler boosting as tools to measure the secondary mass in the next Section.


\section{Results}
\label{ResultsSect}

\subsection{Overview}
\label{MainResultsSect}

Table \ref{VarTable} lists numerical results for the expected numbers of transit detections with $r_{\rm p} < 1.4$ R$_{\rm Jup}$, while Table \ref{Var2Table} details how uncertainties in individual parameters perturb the number of objects detected. Table \ref{MassTable} shows how our detections separate out into the various mass- and radius-based regimes.

A modestly large number of exoplanets should be detectable with this technique. The sample of exoplanets this large is greater than the number of known planets today, including the \emph{Kepler} planet candidates. However, this will not be the case by the time \emph{Euclid} is launched, as other missions such as \emph{TESS} are expected to have discovered many thousands of candidates in the interim. \vis{} detections outweigh the \nisp{}-$H$ detections. $H$-band transits are around objects that would be identifiable in \vis{}: this is a requirement of setting a more-liberal blending limit for \nisp{}-$H$ detections. For all reasonable input parameters, our detections are dominated by planets of Neptune mass and above, orbiting late-F and early-G stars in orbits of 2--10 days. We can expect to detect several thousand objects of planetary radius in \vis{} at $s_{\rm t} > 10$. We can further expect to detect several hundred in \nisp{}-$H$. We can expected these to be the same objects, and the combination of detections in two different bands can give a near-independent confirmation of their detection, plus provide information to rule out false positives (Sections \ref{BlendsSect}--\ref{FalsePulsSect}).

\begin{center}
\begin{table*}
\caption{Expected numbers of detectable transits from planets, sub-stellar objects and low-mass stars with radius $R<1.4 R_{\rm Jup}$ and signal-to-noise $s_{\rm t} > 10$.}
\label{VarTable}
\begin{tabular}{lrrrrl}
    \hline \hline
Model & All \vis{} detections  & those with $b > 0.75$ & All \nisp{}-$H$ detections  & those with $b > 0.50$ & \\
    \hline
Simple model					& 41 782 	   & 24 923		&  4 288	 & 1 928 \\
Best-estimate physical model			&  4 765 	   &  4 067		&    749	 &   630 \\
Physical model, 1$\sigma$ confidence interval	&  3 374 --  8 692 &  2 896 --  6 986   &   505 -- 1 777 & 442 -- 1 552\\
Physical model, 2$\sigma$ confidence interval	&  1 975 -- 15 026 &  1 766 -- 12 246   &   228 -- 5 611 & 195 -- 4 161\\
    \hline
\multicolumn{5}{p{0.97\textwidth}}{Note: the number of detections is normalised to observed values by using $C_{\rm m,p} = 0.105$ in the range $0.3 < m_\ast < 10$ and $2 < p_{\rm o} < 2000$ days following \citet{CBM+08}. The confidence interval of the physical model takes into account uncertainties in the planet frequency and correlated (`red') detector noise as listed in Section \ref{ModDistAdoptSect}, whether grazing transits are counted, and a 10 per cent allowance for uncertainties in the Galactic model.}
\end{tabular}
\end{table*}
\end{center}

\begin{center}
\begin{table*}
\caption{Relative impact of each parameter on the number of detectable ($s_{\rm t} > 10$ in \vis{}), unblended ($b_{VIS,H} > 0.75$, 0.5) planets ($R < 1.4 R_{\rm Jup}$), compared to the best-estimate physical model.}
\label{Var2Table}
\begin{tabular}{lcccccccl}
    \hline \hline
\  & 
\multicolumn{1}{c}{$\alpha_{\rm m}$} &
\multicolumn{1}{c}{$\beta$} &
\multicolumn{1}{c}{$\gamma$} &
\multicolumn{1}{c}{$\delta$} &
\multicolumn{1}{c}{$\log(Q_1)$} &
\multicolumn{1}{c}{Correlated noise} &
\multicolumn{1}{c}{Grazing transits} &
\\
    \hline
Worst-case limit				&   0    &  0.5  & 0.5  &  1.8  &  6.5 & 3000 ppm & Exclude \\
Nominal case					& --0.31 &  0.26 & 1.0  &  2.6  &  7.5 & 1000 ppm & Include \\
Best-case limit					& --0.5  &  0    & 1.5  &  3.5  &  8.5 &  300 ppm & Include \\
    \hline
Worst-case change (per cent, \vis{})		& --12   & --53  & --10 &  --3  & --56 &  --42    &  --2 \\
Best-case change (per cent, \vis{})		&  +9    &  +94  &  +12 &  --1  &  +87 &   +11    &    0 \\
Worst-case change (per cent, \nisp{}-$H$)	& --19   & --60  & --32 &  --12 & --58 &  --56    &  --2 \\
Best-case change (per cent, \nisp{}-$H$)	&   0    &  +95  &  +21 &  --5  &  +91 &    +2    &    0 \\
    \hline
\end{tabular}
\end{table*}
\end{center}



\begin{center}
\begin{table*}
\caption{Unblended objects in the best-estimate physical model. For each radius category, the percentage in each mass classification is given.}
\label{MassTable}
\begin{tabular}{l r@{\ \ }r@{\ \ }r@{\ \ \ \ }r r@{\ \ }r@{\ \ }r@{\ \ \ \ }r}
    \hline \hline
\multicolumn{1}{l}{Group} & \multicolumn{4}{c}{\vis{}-band detections} & \multicolumn{4}{c}{$H$-band detections} \\
\multicolumn{1}{c}{\ } & 
\multicolumn{1}{c}{Pl} & \multicolumn{1}{c}{BD} & \multicolumn{1}{c}{LMS} & \multicolumn{1}{c}{Total} & 
\multicolumn{1}{c}{Pl} & \multicolumn{1}{c}{BD} & \multicolumn{1}{c}{LMS} & \multicolumn{1}{c}{Total} \\
    \hline
Small planets ($R < 0.7$ R$_{\rm Jup}$)			&    100\rlap{\%} &     0\rlap{\%} &     0\rlap{\%} &     92    &    100\rlap{\%} &    0\rlap{\%} &     0\rlap{\%} &      5  \\
Jupiter-radius objects ($0.7 < R < 1.4$ R$_{\rm Jup}$)	&     96\rlap{\%} &     3\rlap{\%} &     1\rlap{\%} &   4050    &     97\rlap{\%} &    2\rlap{\%} &     1\rlap{\%} &    242  \\
Small stars ($R > 1.4$ R$_{\rm Jup}$)			&     72\rlap{\%} &     3\rlap{\%} &    25\rlap{\%} &   1159    &     82\rlap{\%} &    2\rlap{\%} &    16\rlap{\%} &    212  \\
Total							&   4814          &   167          &   321          &   5301    &    413          &    8          &    38          &    458  \\
    \hline
\multicolumn{9}{p{0.73\textwidth}}{Note that rounding errors introduced by the normalisation constant may make some additions not tally. Being photometrically differentiable from planets, the ``small stars'' category is not included in figures quoted in the text or Table \ref{VarTable}. Abbreviations: Pl = planets ($m < 13$ M$_{\rm Jup}$); BD = brown dwafs ($13 < m < 75$ M$_{\rm Jup}$); LMS = low-mass stars ($m > 75$ M$_{\rm Jup}$).}
\end{tabular}
\end{table*}
\end{center}

\subsection{Distribution of parameters}
\label{ParamSect}

\begin{figure}
\centerline{\includegraphics[height=0.50\textwidth,angle=-90]{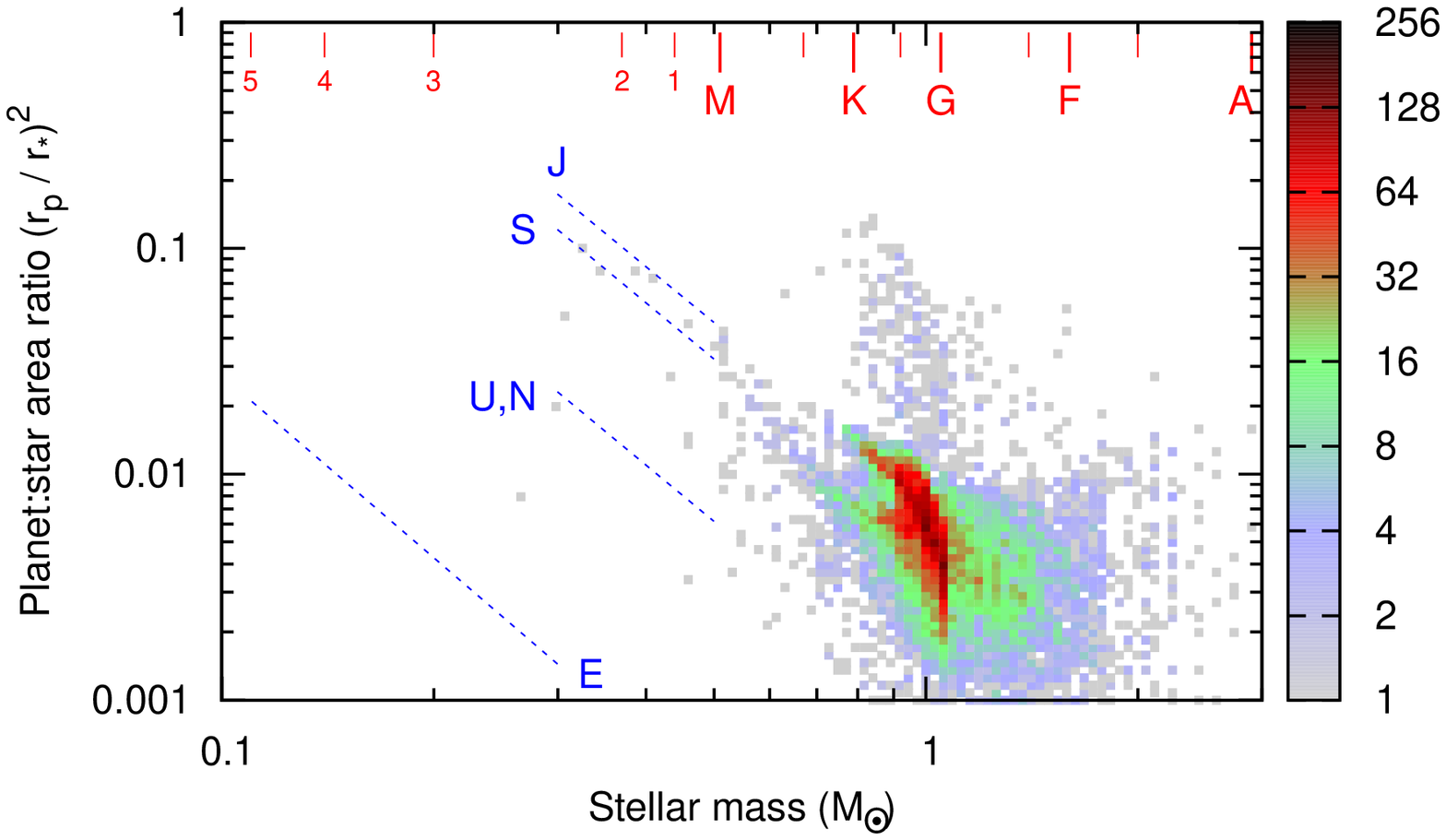}}
\caption{Host mass versus transit depth (as defined by the ratio of areas of the secondary to primary objects) for `unblended' stars with transits detected at $>$10$\sigma$ in \vis{}. Vertical and diagonal lines denote \emph{approximate} spectral classes for Population II main sequence stars and Solar System planets, respectively.}
\label{MSRPFig}
\end{figure}

\begin{figure}
\centerline{\includegraphics[height=0.50\textwidth,angle=-90]{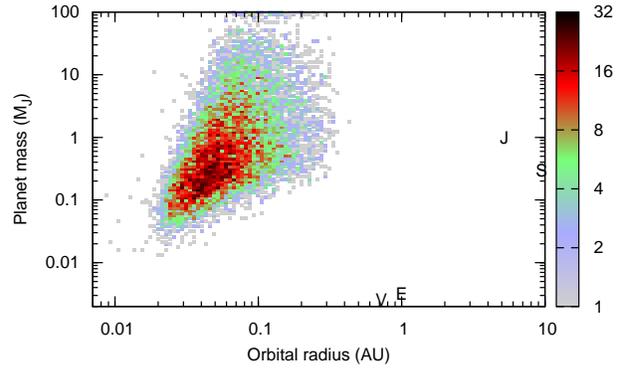}}
\caption{Simulated detections of companion mass and orbital radius.}
\label{MPROFig}
\end{figure}

\begin{figure}
\centerline{\includegraphics[height=0.50\textwidth,angle=-90]{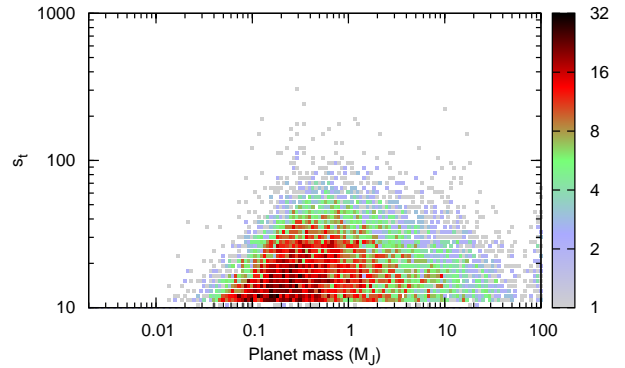}}
\caption{Signal-to-noise of transit detection as a function of companion mass. Our applied sensitivity limit is $s_{\rm t} = 10$.}
\label{SigmaTFig}
\end{figure}

\begin{figure}
\centerline{\includegraphics[height=0.50\textwidth,angle=-90]{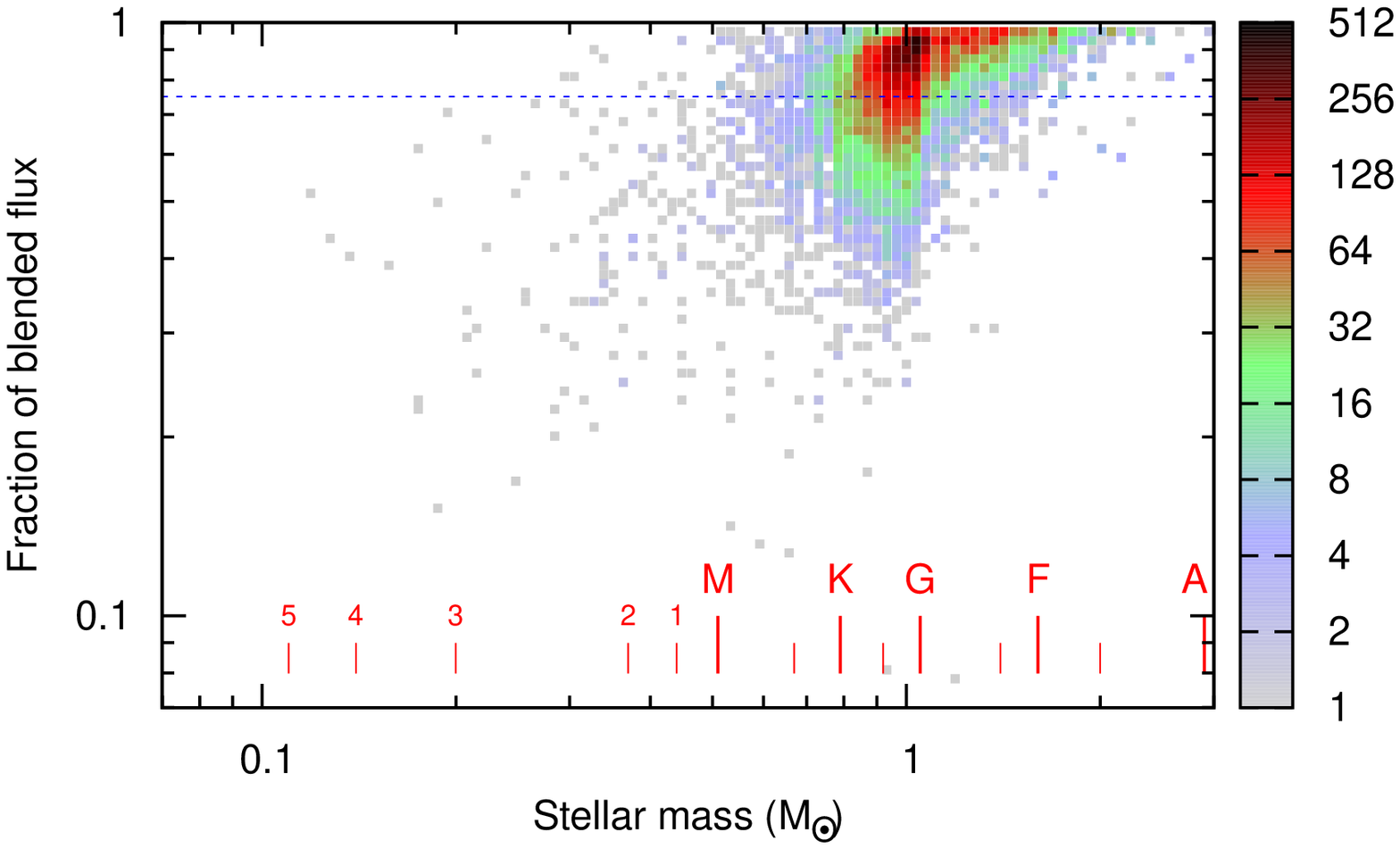}}
\centerline{\includegraphics[height=0.50\textwidth,angle=-90]{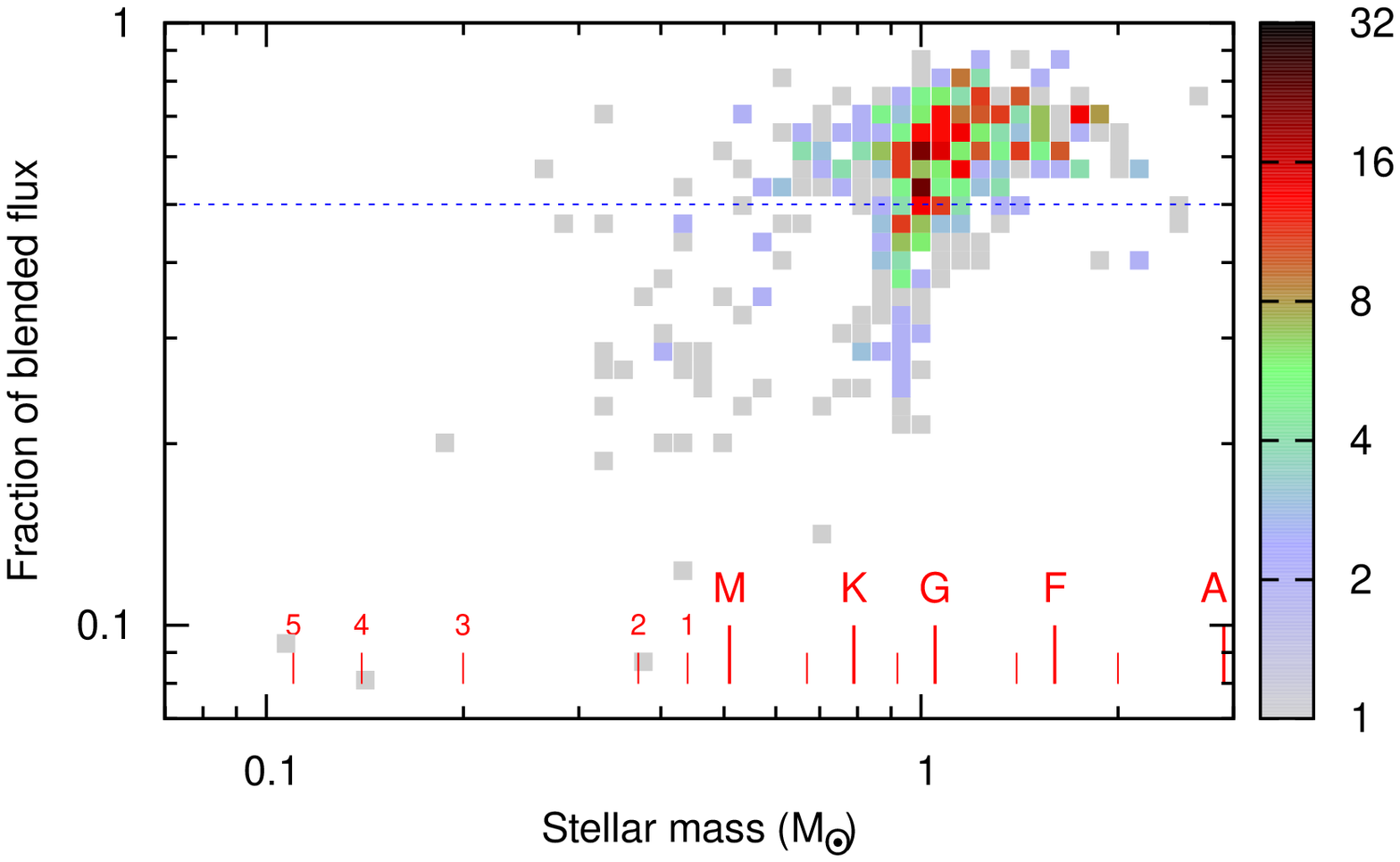}}
\caption{Fraction of the flux in the photometric aperture arising from the primary host star, $b$, as a function of stellar mass. Vertical lines represent approximate spectral classes for main sequence stars. Only stars above the horizontal line are considered in the analysis and shown in the other Figures. The top panel shows \vis{} detections, the bottom panel shows $H$-band detections.}
\label{BlendFig}
\end{figure}

\begin{figure}
\centerline{\includegraphics[height=0.50\textwidth,angle=-90]{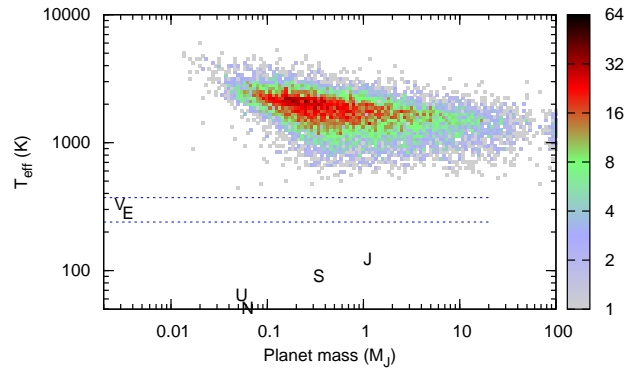}}
\caption{Companion mass versus companion equilibrium temperature. An approximate habitable zone (defined by 240--372 K) is marked, along with the positions of Solar System planets for albedo $A = 0$.}
\label{HZFig}
\end{figure}

Figures \ref{MSRPFig} through \ref{HZFig} show the distribution of parameters we expect for our best-estimate model.

Figure \ref{MSRPFig} shows the mass of the host stars of our detectable sub-stellar companions. We expect detections from early A to around M0. However, the majority of \vis{} detections are around late-F and G-type stars. These stars are typically in the Galactic Bulge. Later-type stars with detectable companions are closer to us ($d < 4$ kpc) in the Galactic Disc. While the $H$-band detections do include Bulge stars, they favour the closer stars in the Galactic Disc.

Figure \ref{MPROFig} shows the distribution of companion masses and orbital radii. The typical \vis{} detection is of a 0.9-M$_{\rm Jup}$ hot Jupiter orbiting at 0.05 AU in 4.2 days. The typical $H$-band detection is a 1.0-M$_{\rm Jup}$ hot Jupiter orbiting at 0.05 AU in 3.9 days. We reach our sensitivity limit close to Neptune-mass planets, and we do not expect to reach the required sensitivity to detect rocky planets.

Figure \ref{SigmaTFig} shows the signal-to-noise in the folded transit lightcurves. A very few high-quality (S/N $>$ 100) transits are expected in \vis{}, but most will be much closer to our signal-to-noise cutoff. The same is true at $H$-band. Determining whether the transit is `grey' (i.e.\ the transit depth is the same in both bands) can be a useful determinant in resolving blended eclipsing binaries (Section \ref{FalseBEBsSect}). Differences between $H$-band and \vis{} transit depth could be used to probe the wavelength-dependence of the diameter of companions (e.g.\ \citealt{PKG+08}). However, a $\sim$1 per cent change in transit depth would require $s_{\rm t} > 100$ in both \vis{} and $H$-band, meaning such transmission spectrophotometry of companion atmospheres is not likely to be possible. It would be possible, however, to use the combination of \vis{} and $H$-band to increase the signal-to-noise of a candidate detection. 

Figure \ref{BlendFig} shows the fraction of light arising from blends. Our (conservative) cuts of $b > 0.75$ in \vis{} and $b > 0.50$ in $H$-band has meant we retain most of the detectable transits. The lower value in $H$-band is a requirement if we are to use these bands to characterise the transits. Blending means we lose planets around the lowest-mass stars, and extracting there will depend on our ability to measure the host magnitudes in the presence of nearby stars, which may require ground-based adaptive-optics follow-up.

Figure \ref{HZFig} shows the equilibrium (zero albedo, phase-averaged) temperatures of the companions discovered using \vis{}. While we theoretically have the sensitivity to detect both Earth-sized planets and planets in the habitable zone (240--372 K; Figure \ref{ContourFig}), it is quite unlikely that we will observe these due to the necessity for the planet to have a favourably-inclined orbit, with transits matching our observing windows, around a star with high signal-to-noise and low blending.

We therefore conclude that \euclid{} can easily detect a large number of hot Jupiters around resolved stars in the Galactic Bulge. However, at 1.2 m, \euclid{}'s design is the smallest effective system for this work. A smaller telescope would lack both the sensitivity to detect most of the transiting companions in this photon-noise-limited regime, and the resolution to separate host stars from nearby blends. While \euclid{} may lack \kepler{}'s sensitivity to low-mass, rocky planets, we can expect the number of transiting (sub-)Jupiter-radius objects found by \euclid{} to be the same order of magnitude as those found by \kepler{}.

\subsection{Variation of the distribution}
\label{UncertainSect}

Table \ref{Var2Table} shows the differences in detection numbers that result from taking our best-estimate physical model and varying each parameter in turn. Correlations between parameters such as $\beta$ and $Q_1$ mean that these cannot be simply added to create the total uncertainty, which is better modelled in our Monte-Carlo trials, summarised in Table \ref{VarTable}. In the remainder of this Section, we describe how adjusting each parameter affects the number and distribution of sub-stellar objects we find.

The similarities of spectral types of host stars in \exels{} and \emph{Kepler}, coupled with the adoption of the normalisation constant, $C_{\rm m,p}$, mean that several parameters have little overall impact on the distribution. For example, exploring the plausible range for the planet and stellar mass exponents ($\alpha_{\rm m}$ and $\gamma$) changes the number of detectable transits by $<$30 per cent. The same is true of the metallicity exponent, $\delta$, though we remind the reader of the additional few percent change due to the difference between the \besancon\ and observed metallicity distribution (Section \ref{ModDistDistSect}). Grazing transits represent a small fraction of the total population (1--2 per cent). Whether we are able to extract the transit signatures of grazing companions does not have a great impact on the number of sub-stellar companions we detect.

\begin{figure}
\centerline{\includegraphics[height=0.50\textwidth,angle=-90]{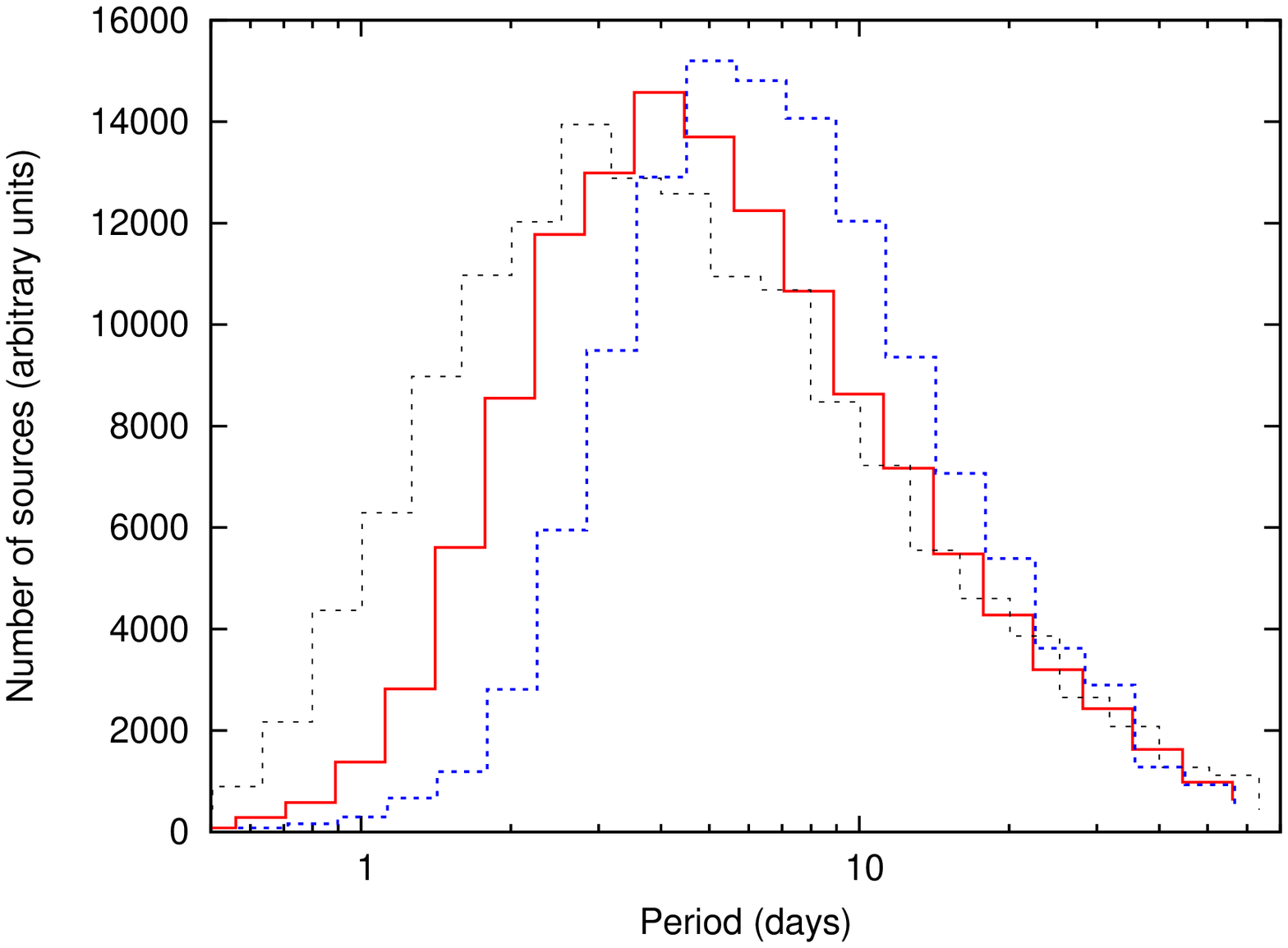}}
\centerline{\includegraphics[height=0.50\textwidth,angle=-90]{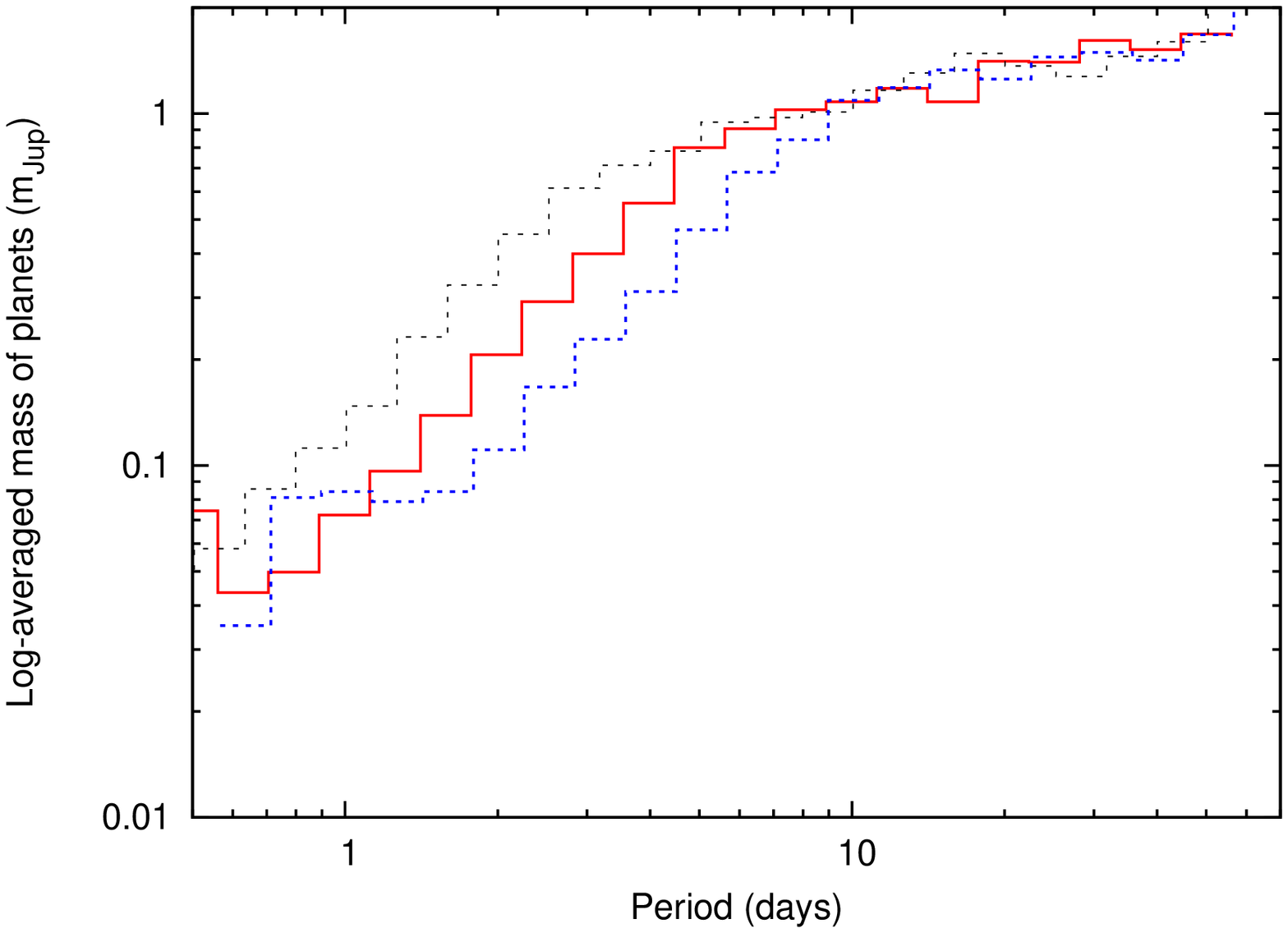}}
\caption{The distribution of orbital periods for differing values of $\beta$ and $Q_1$ (top panel) and the average mass as a function of orbital period (for $\alpha = -0.31$). The solid, red lines gives our best-fit model: $\beta = 0.26$, $\log Q_1 = 7.5$. The black, dotted lines shows results for $\beta = 0$, $\log Q_1 = 8.5$. The blue, dashed lines shows results for $\beta = 0.5$, $\log Q_1 = 6.5$.}
\label{BetaQFig}
\end{figure}

The orbital period exponent, $\beta$, and the tidal decay parameter, $Q_1$, operate together to shape the orbital period distribution of exoplanets. Individually, these parameters can change the total number of exoplanets we will observe by a factor of two in either direction, but our normalisation constant means they will anti-correlate such that their combined effect on the total number cannot be much more than a factor of two. Figure \ref{BetaQFig} shows that the peak of the period distribution can provide information about the strength of the tidal connection between star and planet. As the Bulge is an old population (e.g.\ \citealt{GZHS14}), this effect may be stronger than the Solar Neighbourhood. However, we note that it may be difficult to extract this information from variations in the initial planet distribution (other than $\beta$). Some clue may come from its imprint in the average planet mass (Figure \ref{BetaQFig}; lower panel), though this effect could also be mimicked by planetary evaporation.

Correlated noise has very little effect on the number of companions we expect to detect at \vis{} until it exceeds $\sim$2000 ppm in a transit. In this regime, our noise sources are dominated by the Poisson noise of the incoming photons. The limiting value of red noise in $H$-band is less, at $\sim$1000 ppm, as blending is higher and correlated noise from blended stars starts to become important. Unfortunately, while they may end up being the most scientifically interesting, additional correlated noise affects the smallest planets around the smallest stars more. We may expect a decreased number of these if correlated noise cannot be sufficiently reduced. However, the reduction of correlated noise for transit detection is a well-established science, and whitening of data to the level of tens of ppm is easily achievable for satellites such as \emph{Kepler}. Provided the correlated noise does not have a strong impact at frequencies comparable to the transit duration, we can expect it to be largely removable. If all red noise can be removed, the number of transit detections would increase by 13 per cent from the best-estimate model.

\begin{figure}
\centerline{\includegraphics[height=0.50\textwidth,angle=-90]{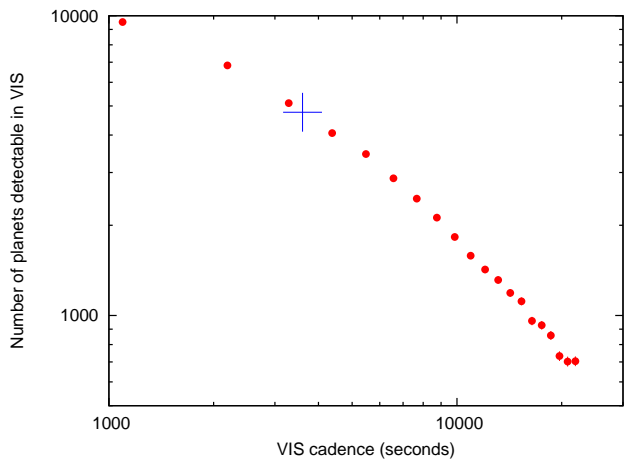}}
\caption{Number of sources detected in our best-estimate physical model for different \vis{} cadences. The plus sign marks our nominal, hourly cadence.}
\label{CadFig}
\end{figure}

The cadence of the \vis{} observations has a significant effect on the number of companions we detect. We have assumed a \vis{} cadence of once per hour in our models, imposed by the download limit from the spacecraft. Figure \ref{CadFig} shows the number of expected objects detected with $R < 1.4$ R$_{\rm Jup}$ for a variety of different cadences. Increasing the \vis{} cadence to once per 1095 seconds would approximately double the number of detectable planets. Conversely, reducing the cadence to once per 12 hours (as per $Y$- and $J$-band) means a transit survey becomes unviable. Nevertheless, we may still expect transits of $\approx$1 per cent of objects detected in \vis{} to be detectable in the 12-hourly $Y$ and $J$-band exposures too.

\begin{figure}
\centerline{\includegraphics[height=0.50\textwidth,angle=-90]{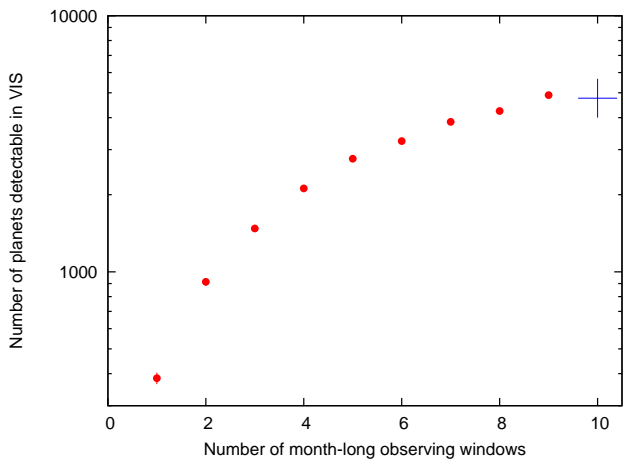}}
\caption{Number of sources detected in our best-estimate physical model for different numbers of observing windows. Each window is assumed to last 30 days and windows are spaced six months apart. The plus sign marks our nominal 10-month survey.}
\label{EpochsFig}
\end{figure}

The total observing time is also modelled (Figure \ref{EpochsFig}). This survey assumes that we observe in ten month-long periods, spaced every six months. The number of detected planets increases close to linearly with observing time with observing times greater than $\approx$4 months. Below this, only the closest-in planets around the brightest stars will be detected. We note that a long observing period is key to picking up planets at larger orbital radii, which is important for an accurate comparison between the microlensing and transiting planet populations.

Our ability to perform photometry on more-heavily-blended stars from the images can make a crucial difference to the number of companions we can detect. Our default assumptions ($b > 0.75$ for \vis{} and $b > 0.50$ for $H$-band) are set such that we expect to initially identify both individual stars and their transiting companions in the cleaner, higher-resolution \vis{} images, then use this information to identify those same stars and companions in the $H$-band data. This is vital to the success of detecting companions at $H$-band, as only a fairly small fraction of stars will have $b > 0.75$ in the $H$-band images (see next Section). A limit of $b > 0.75$ is quite conservative: difference imaging is capable of detecting variability in stars much more blended than this (see, e.g., \citealt{AAA+99}), though its ability to achieve the required photometric accuracy to detect transits of heavily-blended stars in \euclid{} data has not yet explicitly been shown.

\begin{figure}
\centerline{\includegraphics[height=0.50\textwidth,angle=-90]{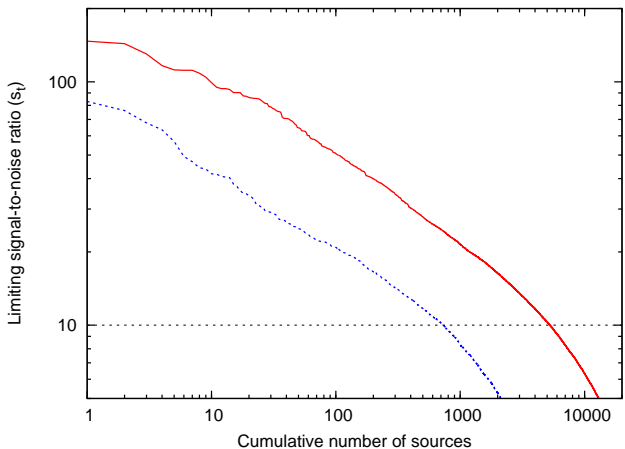}}
\caption{Number of sources detected in our best-estimate physical model versus their signal-to-noise ratio. \vis{} detections are shown in as a red, solid line; \nisp{}-$H$-band detections as a blue, dashed line. Our applied cutoff is shown as the horizontal line.}
\label{SNRFig}
\end{figure}

The minimum signal-to-noise at which we can detect a transit is also a major determinant of the number of companions we can expect to find. Our initial value of 10$\sigma$ is conservative: decreasing this to the 8$\sigma$ of \citet{BKB+08} or the 6$\sigma$ of \citet{KZM02} would approximately double the number of detectable companions we may hope to detect (Figure \ref{SNRFig}).

Our final detection numbers are therefore mostly limited by the success of the reduction pipeline in detecting lower-signal-to-noise transits around heavily-blended stars. For the currently-envisaged survey strategy corresponding to our best-estimate model, our limits are set by photon noise in \vis{}, which means high-cadence observations in the \vis{} channel are vital to detecting large numbers of companions. As we have chosen conservative estimates for our model parameters, we can expect the true number of observable transiting companions will be considerably greater than those listed in Table \ref{VarTable}.

\subsection{Comparison with microlensing detections}
\label{MicroSect}

\begin{figure*}
\centerline{\includegraphics[height=0.78\textwidth,angle=-90]{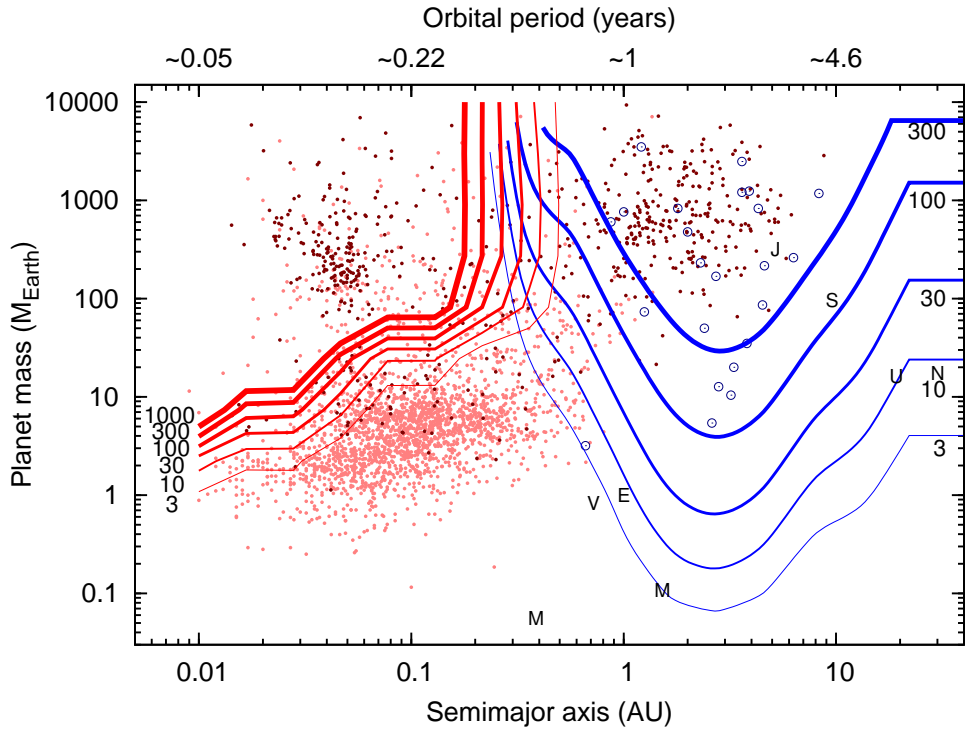}}
\caption{Detection sensitivity, showing the number of planets detected at $s_{\rm t} > 10$ if each star has a planet at that particular mass and radius (contoured in red, normalised to $2.9 \times 10^7$ stars). Contoured in blue are the expected microlensing events presented in Paper~I. Dots show known exoplanets (red) or exoplanet candidates (pink), with planets detected through microlensing circled in blue. Data from: http://www.exoplanets.org.}
\label{ContourFig}
\end{figure*}

\begin{figure*}
\centerline{\includegraphics[height=0.78\textwidth,angle=-90]{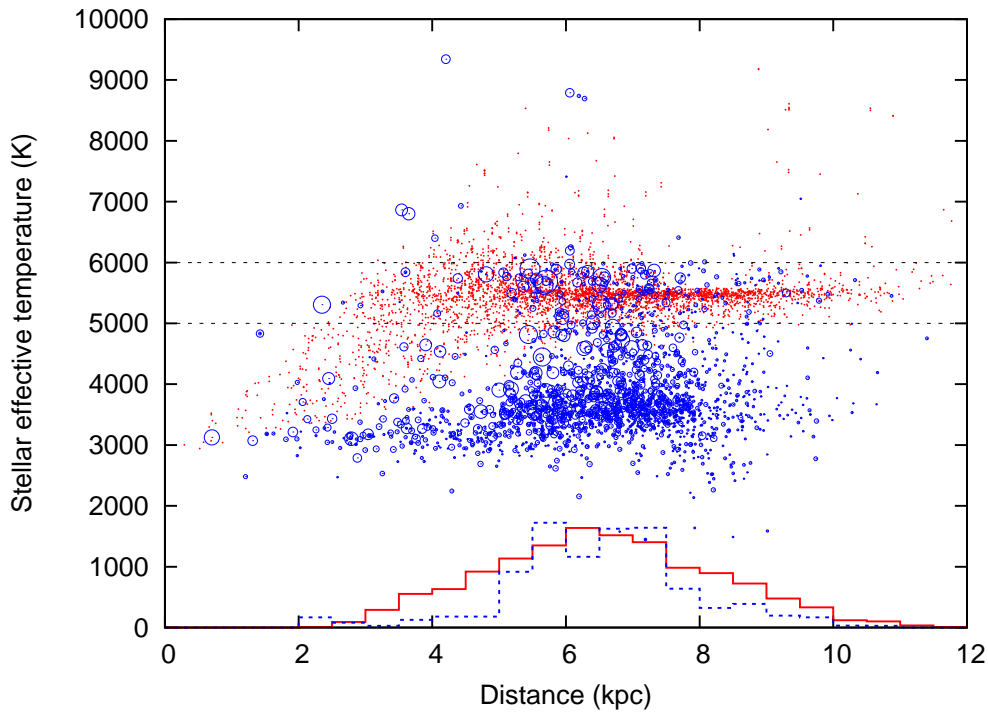}}
\caption{Stellar effective temperature and distance for transiting exoplanets corresponding to the best-estimate model of this work (red dots) compared to the microlensed planets of Paper~I (symbol area scales linearly with detection probability). The horizontal lines denote the temperature region ($\sim$F and G stars) over which the distance distribution of microlensed and transiting exoplanets are similar enough to provide a comparable sample. The distance distributions of stars in this temperature range are shown in the histograms at the bottom of the plot (red, solid and blue, dashed lines for transits and microlensing events, respectively): magnitude limits could also be imposed to confine the distance range when comparing the frequency of microlensed to transiting planets. Note that the lack of microlensing events below $\sim$3000 K is an artifact imposed by the lower mass limit of the \besancon{} models.}
\label{MicroDistFig}
\end{figure*}

Figure~\ref{ContourFig} shows the sensitivity of both the \exels{} transit survey and the microlensing survey discussed in Paper~I. While it is clear that the transit survey will not approach the sensitivity of \kepler{}, its sensitivity extends beyond that of most currently-published ground-based transit surveys. Temporal coverage means sensitivity declines towards the orbital radius of Mercury, while photon noise limits our sensitivity for planets below Neptune's mass.

Importantly, the coverage at super-Jupiter masses is almost continuous between the regime where the transit survey and microlensing survey are both sensitive. We will therefore have near-continuous coverage in orbital radius from the closest exoplanets to those at tens of AU from their host star.

Figure~\ref{MicroDistFig} shows the comparative distances and effective temperatures of the microlensing and transiting exoplanet hosts. While the microlensed planet hosts are typically around much cooler, smaller stars, the distance distributions of microlensed and transiting planets among late-F- and G-type stars shows roughly equal distributions of stars in the Galactic Disk ($<$6 kpc) and Galactic Bulge ($>$6 kpc), in which the Galactic Bulge targets dominate both samples due to the location of the Bulge main-sequence turn-off. Further discrimination of Bulge versus Disk targets may come from source proper motions and apparent brightness (plotted as a colour--magnitude diagram). This confirms that we can make a direct calibration between the detection efficiency of the transit survey and that of the microlensing survey, allowing us to constrain the variation in planet frequency with orbital radius (parametrised by $\beta$ in Eq.\ (\ref{PlDensityEq4})) across, effectively, the entire range of orbital radii from sub-day orbits to the free-floating planet regime.

\subsection{Distinguishing planets from brown dwarfs}
\label{DayNightSect}

\subsubsection{Object size}
\label{DSizeSect}

Table \ref{MassTable} lists the various fractions of planets, brown dwarfs and low-mass stars in our small planets, Jupiter-radius objects and small stars regimes. In all cases, our observational setup ensures that all objects below our $r_{\rm p} = 0.7$ R$_{\rm Jup}$ cut-off are genuine small planets. This is consistent with the lack of observed transiting brown dwarfs with radii below 0.7 R$_{\rm Jup}$ (Figure \ref{MRFig}). However, there are only likely to be few planets that we can confidently claim are less than 0.7 R$_{\rm Jup}$ in all but the most-favourable observing scenarios.

Statistically, however, most Jupiter-radius objects we observe should be \emph{bona fide} planets. Simplistically, the mass corresponding to Jupiter-radius planets ($\approx$0.1 to 13 M$_{\rm Jup}$) is large ($\approx$2.1 dex) compared to that covered by Jupiter-radius brown dwarfs (13 to $\approx$160 M$_{\rm Jup}$, $\approx$1.1 dex), leading to a preference for detecting planets. When coupled with the preferential tidal decay of orbits of massive companions, and a negative $\alpha_{\rm m}$, this leads to a strong selection bias favouring the detection of planets over brown dwarfs and low-mass stars. Any Jupiter-radius object we find therefore has a $\approx$96 per cent chance of being a gas-giant planet, before any other factors are considered. We note, however, that this statistic does not include the false positive rate of objects mimicking transits.

Conversely, we also find that most of the objects we define as small stars (based on their radius) are actually heavily-bloated planets, with dwarf stars representing only $\sim$10--30 per cent of objects. We typically probe systems with ages of $>$5 Gyr, hence any bloating is expected to be purely a thermal equilibrium effect due to absorption of light from the host star, rather than a remnant of their primordial collapse.

\subsubsection{Doppler boosting and ellipsoidal variation}
\label{DBandESect}

For our best-estimate simulations, the maximum amplitude of the Doppler boosting and ellipsoidal variation signals we could expect for small stars are both $\sim$100--200 $\mu$mag, exceptionally up to $\sim$300 $\mu$mag. It is unlikely that this could be detected with high confidence, even if red noise can be mitigated above our expectations. However, in some cases, it may provide a useful limit to the maximum mass of the planetary body and differentiate planetary-mass bodies from very-low-mass stars and brown dwarfs, which should produce signals a few times higher, which may be detectable in some cases.

Our simulations also do not take into account any rare, short-lived objects that are gravitationally scattered into orbits that are unstable on the systemic timescale. Such systems tend to have high orbital inclinations and non-zero eccentricities. Hence, while we predict that ellipsoidal variations in these cases may reach $\sim$1 mmag, Eq.\ \ref{EllipseEq} is likely a poor representation of the ellipsoidal variation we would achieve in these cases. Instead, we may either see variations in the transit lightcurve due to gravity darkening (cf.\ KOI-13.01; \citealt{BLS11}) or tidally-excited pulsations on the stellar surface (cf.\ \citealt{HMRN11,HKP+13}), both of which can produce effects with $\sim$mmag amplitudes.

\subsubsection{Emission and reflection detection}
\label{EandRSect}

\begin{center}
\begin{table}
\caption{Expected detections of planetary emission and reflection for secondary eclipses and orbital modulation, based on our best-estimate model.}
\label{ERTable}
\begin{tabular}{llrrrr}
    \hline \hline
Albedo & $\epsilon^{1}$ & \multicolumn{2}{c}{\vis{} sensitivity} & \multicolumn{2}{c}{$H$-band sensitivity} \\
\  & \  & \multicolumn{1}{c}{$>$1$\sigma$} & \multicolumn{1}{c}{$>$3$\sigma$} & \multicolumn{1}{c}{$>$1$\sigma$} & \multicolumn{1}{c}{$>$3$\sigma$} \\
    \hline
\multicolumn{6}{c}{\emph{Secondary eclipse detections}} \\
1.0	&  0\%	& 313	& 46 	& 35 	& 16 	 	\\
1.0	& 50\%	& 212 	& 19 	& 31 	& 3 	 	\\
1.0	& 100\%	& 144	& 7 	& 12 	& 0 	 	\\
0.5	& 50\%	& 157 	& 11  	& 25	& 2		\\
0.5	& 100\%	& 81 	& 3  	& 11	& 1		\\
0.2	& 33\%	& 132 	& 9  	& 21	& 7		\\
0.0	& 50\%	& 81	& 3	& 20	& 2	 	\\
0.0	& 100\%	& 34	& 2	& 11	& 6		\\
\hline
\multicolumn{6}{c}{\emph{Orbital modulation detections}} \\
1.0	&  0\%	& 221	& 29	& 52 	& 1 	 	\\
1.0	& 50\%	& 182	& 12 	& 13 	& 0 	 	\\
1.0	& 100\%	& 91	& 14 	& 2 	& 0 	 	\\
0.5	& 50\%	& 164	& 4 	& 5	& 0		\\
0.5	& 100\%	& 96 	& 0	& 0	& 0		\\
0.2	& 33\%	& 144 	& 23  	& 10	& 0		\\
0.0	& 50\%	& 98	& 6	& 0	& 0		\\
0.0	& 100\%	& 58	& 1	& 0	& 0		\\
     \hline
\multicolumn{6}{l}{$^1$Heat transport efficiency from day- to night-side.}
\end{tabular}
\end{table}
\end{center}




The literature suggests albedoes of $<$20 per cent and heat redistributions of $\sim$33 per cent may be relatively typical for hot Jupiters \citep{Barman08,RMS+08,CA11,KS11,DSM+11,KB11,SAS+11}. Planets are therefore typically much blacker than those in our Solar System (cf.\ \citealt{Allen00}). Table \ref{ERTable} lists the expected number of detections of secondary eclipses and orbital modulations that we can expect to detect for our best-estimate model, for a variety of planetary albedoes and day-night heat transport efficiencies. Obviously, objects with 1--3$\sigma$ detections will not be identifiable by themselves. However, in a statistical context they could be used to place constraints on the albedos of transiting exoplanets. With these parameters, and our expected sensitivity and observing setup, we should collect $\sim$130 secondary eclipses but only $\sim$10 orbital modulation observations at 1$\sigma$, with only $\sim$9 secondary eclipses actually making a $>$3$\sigma$ detection. We should therefore be able to make a statistical estimate of the average planetary albedo, but are not likely to greatly constrain the heat transport efficiency. We are limited here by photon noise, so the number of detections should not vary with reasonable amplitudes of red noise in the lightcurves.

\subsubsection{Follow-up photometry and spectroscopy}
\label{FollowSpectSect}

\begin{figure}
\centerline{\includegraphics[height=0.50\textwidth,angle=-90]{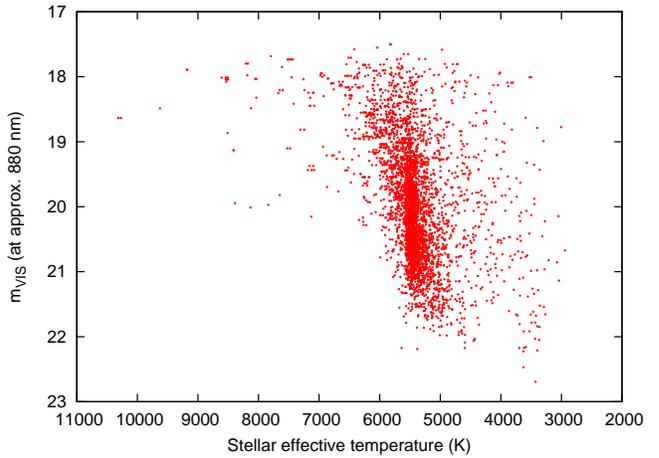}}
\caption{Temperature--magnitude diagram for the stars with transiting companions ($R < 1.4$ R$_{\rm Jup}$) detectable using \vis{}.}
\label{CMDFig}
\end{figure}

Spectroscopic follow-up is a bottleneck for confirming planets among many different surveys (e.g.\ \citealt{Ricker14}). Our brightest candidates are around relatively unreddened stars of $\sim$17th magnitude ($\vis{} \approx RI_cZ = 17.5$ mag; $Y \approx 17$ mag (Figure \ref{CMDFig}), hence our survey will be no exception. As examples, 3-hour integrations under good conditions on a star of $R = 18$ with the ESO-3.6m/HARPS spectrograph would attain signal-to-noise ratios of $\approx$3 in the 6000 \AA\ region commonly used for radial velocity measurements, while similar observations with the ESO-VLT/UVES spectrograph would attain a signal-to-noise ratio of $\approx$23. The next generation of telescopes may fair better, with the 39-m Extremely Large Telescope obtaining a signal-to-noise roughly five times greater. A UVES-like instrument could then achieve a signal-to-noise of $\sim$110 over the period of transit, or $\sim$36 in a one-hour integration. This may allow radial velocity confirmation for a selection of compelling objects. Of particular interest may be the Rossiter--McLaughlin effect, through which orbital--rotational alignments can also be explored (cf.\ \citealt{GW07}), though signal-to-noise limits may mean several transits would need to be observed to make a significant detection.

Spectral characterisation of a wide variety of host stars should also be possible, with which one can explore the planet occurrence with host composition. Accurate metallicities and abundances will be possible for many of the targets in our sample. It is relatively trivial to get accurate stellar abundances down to at least $m_{\rm VIS} \approx 19$ with an 8-m-class telescope (e.g.\ \citealt{HKW+14}). Use of a near-infrared spectrograph like VLT/MOONS\footnote{Multi-Object Optical and Near-Infrared Spectrograph; http://www.roe.ac.uk/$\sim$ciras/MOONS/Overview.html; \citet{ODG+12}.} may permit characterisation of cooler and/or heavily reddened stars (e.g.\ \citealt{LNH+14}).

Photometric follow-up of objects may be more easily performed. A signal-to-noise of $\sim$100, required to detect transits, would be attainable on moderately-faint candidates even with a 4-m-class telescope with adaptive optics under good conditions. Follow-up photometric confirmation is likely to be limited to transit timing variations. As such, it would mainly be useful in investigating cases where the planet's orbit is being perturbed by a third object, or follow-up of systems evolving on very short timescales. Instruments such as ESO-VLT/FORS can theoretically reach a signal-to-noise of $\sim$1000 on an $I = 20$ mag star over the course of a transit time: sufficient for multi-wavelength measurements of the planet's atmosphere at secondary eclipse. However, space-based photometry would be more likely to actually obtain this quality of data due to atmospheric effects and crowding, ideally with an instruments such as the \emph{James Webb Space Telescope}.

\section{Conclusions}
\label{ConcSect}

We have modelled the expected number of low-mass, transiting companions detectable by \exels{}, a proposed \euclid{} planetary microlensing survey during a ten-month campaign over five years in a 1.64 square degree field near the Galactic Centre. Sensible ranges of input parameters produce several thousand detections and, using typical values from the literature and conservative estimates of mission-specific parameters, we estimate $\sim$4\,000 companions will be detectable, of which the great majority ($\gtrsim$90 per cent) will be hot Jupiters. Physical uncertainties impart a range of error between $\sim$2\,900 and $\sim$7\,000 objects. We expect four-colour photometry (\vis{}, $Y$, $J$ and $H$) for $\sim$20--100 stars and two-colour photometry (\vis{} and $H$) for $\sim$400-1\,600 stars. While this may not provide data on the planets, this will be useful in discriminating stellar blends.

Absolute discrimination of planetary-mass objects from all transiting companions appears prohibitively difficult for all but a few objects. Tidal and relativistic effects will only be detectable for planets scattered into short orbits around their hosts on timescales much shorter than the tidal inspiral time. Given a list of planetary candidates, ground-based confirmation will likely only be successful on new 30-metre-class telescopes.

The separation of planets from brown dwarfs and low-mass stars is probably best performed in a statistical context. Although we do not fully model false-positive detections, the vast majority of the objects we detect in any scenario we have created are planets. The high angular resolution of \euclid{}'s \vis{} instrument should make it relatively easy to distinguish contaminating blends of variable stars from transiting companions, and the clean, frequent sampling should allow us to rule out many false positive categories.

However, \euclid{} will be observing a fundamentally different parameter space to previous space-based surveys. Some 70 per cent of our detections are expected to be around stars in the Galactic Bulge which are older than our Sun, meaning such a survey could discover some of the most ancient planets in the Universe. Their detections therefore have implications for the historical frequency and long-term survival of planetary systems, thus the past habitability of our Galaxy. We may be able to use this advanced age to obtain the strength of the tidal quality factor, $Q_1$, by comparing the period distribution planets to those in the Solar Neighbourhood.

Despite their age, these stars \emph{average} [Fe/H] $\approx$ 0 dex, meaning the formation mechanisms for planets are likely to be broadly similar. However, the large \emph{range} in metallicities means that we can probe how planet abundance changes with host metallicity. Spectroscopic metallicity determination should be easy for at least the brighter objects in our sample, which will also tend to be our best planet candidates due to their higher signal-to-noise ratios.

Most importantly, \exels{} can provide a direct comparison between the frequency of close-in (transiting) and distant (microlensed) Jupiter-like planets, free from major sources of systematic bias, for the first time. While we are limited by the mass--radius degeneracy of Jupiter-radius objects, a robust comparison will be possible for objects smaller than a non-degenerate point further up the mass--radius relationship, e.g. near 2 R$_{\rm Jup}$ and 160 M$_{\rm Jup}$. This should allow the frequency of cold Jupiters to be empirically tied to that of hot Jupiters for the first time. This combination of a transiting and microlensed low-mass companion survey in the Galactic Bulge has the potential to significantly increase our understanding of the frequency and characteristics of low-mass companions in an environment far removed from the Solar Neighbourhood.


\section*{Acknowledgements}

We thank the anonymous referee for their helpful comments which improved the quality and legibility of our paper.



\label{lastpage}

\end{document}